\documentclass[a4paper,12pt, epsfig]{article}
\usepackage{epsfig}
\usepackage{amssymb,latexsym}
\usepackage{amsfonts}
\usepackage{amsmath}
\usepackage[colorlinks, linkcolor=blue, citecolor=blue, urlcolor=blue]{hyperref}
\usepackage{epstopdf}

\newskip\humongous \humongous=0pt plus 1000pt minus 1000pt

\newif\ifdtup

\def\hspp{,\hspace{.5cm}}
\def\hsppp{,\hspace{.25cm}}

\catcode`\@=11

\@addtoreset{equation}{section}
\def\theequation{\arabic{section}.\arabic{equation}}

\def\@normalsize{\@setsize\normalsize{15pt}\xiipt\@xiipt
\abovedisplayskip 14pt plus3pt minus3pt%
\belowdisplayskip \abovedisplayskip
\abovedisplayshortskip \z@ plus3pt%
\belowdisplayshortskip 7pt plus3.5pt minus0pt}

\def\small{\@setsize\small{13.6pt}\xipt\@xipt
\abovedisplayskip 13pt plus3pt minus3pt%
\belowdisplayskip \abovedisplayskip
\abovedisplayshortskip \z@ plus3pt%
\belowdisplayshortskip 7pt plus3.5pt minus0pt
\def\@listi{\parsep 4.5pt plus 2pt minus 1pt
      \itemsep \parsep
      \topsep 9pt plus 3pt minus 3pt}}

\relax

\catcode`@=12

\catcode`\@=11

\def\section{\@startsection{section}{1}{\z@}{3.5ex plus 1ex minus
    .2ex}{2.3ex plus .2ex}{\large\bf}}

\def\thesection{\arabic{section}}
\def\thesubsection{\arabic{section}.\arabic{subsection}}

\def\appendix{\setcounter{section}{0}
  \def\thesection{Appendix \Alph{section}}
  \def\thesubsection{\Alph{section}.\arabic{subsection}}
  \def\theequation{\Alph{section}.\arabic{equation}}}

\def\SymBoxes#1#2#3#4{\newdimen\un@t \un@t#3%
\raisebox{#1}{\rule{#2\un@t}{#4}\hskip-#2\un@t
\@tempdimb\un@t \advance\@tempdimb by-#4\@tempcntb#2\relax%
\@whilenum{\@tempcntb>0}\do{
\rule{#4}{\un@t}\hskip\@tempdimb \advance\@tempcntb by\m@ne}%
\hskip-#2\un@t \rule[\un@t]{#2\un@t}{#4}%
\rule[\un@t]{#4}{#4}\hskip-#4
\rule{#4}{\un@t}}\hskip-#4}                

\begin{document}


\newcommand{\dd}{\textrm{d}}

\newcommand{\beq}{\begin{equation}}
\newcommand{\eeq}{\end{equation}}
\newcommand{\bea}{\begin{eqnarray}}
\newcommand{\eea}{\end{eqnarray}}
\newcommand{\beas}{\begin{eqnarray*}}
\newcommand{\eeas}{\end{eqnarray*}}
\newcommand{\defi}{\stackrel{\rm def}{=}}
\newcommand{\non}{\nonumber}
\newcommand{\bquo}{\begin{quote}}
\newcommand{\enqu}{\end{quote}}
\newcommand{\tc}[1]{\textcolor{blue}{#1}}
\renewcommand{\(}{\begin{equation}}
\renewcommand{\)}{\end{equation}}
\def\de{\partial}
\def\Om{\ensuremath{\Omega}}
\def\Tr{ \hbox{\rm Tr}}
\def\rc{ \hbox{$r_{\rm c}$}}
\def\H{ \hbox{\rm H}}
\def\HE{ \hbox{$\rm H^{even}$}}
\def\HO{ \hbox{$\rm H^{odd}$}}
\def\HEO{ \hbox{$\rm H^{even/odd}$}}
\def\HOE{ \hbox{$\rm H^{odd/even}$}}
\def\HHO{ \hbox{$\rm H_H^{odd}$}}
\def\HHEO{ \hbox{$\rm H_H^{even/odd}$}}
\def\HHOE{ \hbox{$\rm H_H^{odd/even}$}}
\def\K{ \hbox{\rm K}}
\def\Im{ \hbox{\rm Im}}
\def\Ker{ \hbox{\rm Ker}}
\def\const{\hbox {\rm const.}}
\def\o{\over}
\def\im{\hbox{\rm Im}}
\def\re{\hbox{\rm Re}}
\def\bra{\langle}\def\ket{\rangle}
\def\Arg{\hbox {\rm Arg}}
\def\exo{\hbox {\rm exp}}
\def\diag{\hbox{\rm diag}}
\def\longvert{{\rule[-2mm]{0.1mm}{7mm}}\,}
\def\a{\alpha}
\def\b{\beta}
\def\e{\epsilon}
\def\l{\lambda}
\def\ol{{\overline{\lambda}}}
\def\ochi{{\overline{\chi}}}
\def\th{\theta}
\def\s{\sigma}
\def\oth{\overline{\theta}}
\def\ad{{\dot{\alpha}}}
\def\bd{{\dot{\beta}}}
\def\oD{\overline{D}}
\def\opsi{\overline{\psi}}
\def\dag{{}^{\dagger}}
\def\tq{{\widetilde q}}
\def\L{{\mathcal{L}}}
\def\p{{}^{\prime}}
\def\W{W}
\def\N{{\cal N}}
\def\hsp{,\hspace{.7cm}}
\def\hspp{,\hspace{.5cm}}
\def\bo{\ensuremath{\hat{b}_1}}
\def\bfo{\ensuremath{\hat{b}_4}}
\def\co{\ensuremath{\hat{c}_1}}
\def\cfo{\ensuremath{\hat{c}_4}}
\def\th#1#2{\ensuremath{\theta_{#1#2}}}
\def\c#1#2{\hbox{\rm cos}(\th#1#2)}
\def\s#1#2{\hbox{\rm sin}(\th#1#2)}
\def\cp#1#2#3{\hbox{\rm cos}^#1(\th#2#3)}
\def\sp#1#2#3{\hbox{\rm sin}^#1(\th#2#3)}
\def\ctp#1#2#3{\hbox{\rm cot}^#1(\th#2#3)}
\def\cpp#1#2#3#4{\hbox{\rm cos}^#1(#2\th#3#4)}
\def\spp#1#2#3#4{\hbox{\rm sin}^#1(#2\th#3#4)}
\def\t#1#2{\hbox{\rm tan}(\th#1#2)}
\def\tp#1#2#3{\hbox{\rm tan}^#1(\th#2#3)}
\def\m#1#2{\ensuremath{\Delta M_{#1#2}^2}}
\def\mn#1#2{\ensuremath{|\Delta M_{#1#2}^2}|}
\def\u#1#2{\ensuremath{{}^{2#1#2}\mathrm{U}}}
\def\pu#1#2{\ensuremath{{}^{2#1#2}\mathrm{Pu}}}
\def\meff{\ensuremath{\Delta M^2_{\rm{eff}}}}
\def\an{\ensuremath{\alpha_n}}
\newcommand{\Z}{\ensuremath{\mathbb Z}}
\newcommand{\R}{\ensuremath{\mathbb R}}
\newcommand{\rp}{\ensuremath{\mathbb {RP}}}
\newcommand{\vac}{\ensuremath{|0\rangle}}
\newcommand{\vact}{\ensuremath{|00\rangle}                    }
\newcommand{\oc}{\ensuremath{\overline{c}}}
\renewcommand{\cos}{\textrm{cos}}
\renewcommand{\sin}{\textrm{sin}}
\renewcommand{\cot}{\textrm{cot}}

\newcommand{\Vol}{\textrm{Vol}}

\newcommand{\half}{\frac{1}{2}}

\def\changed#1{{\bf #1}}

\begin{titlepage}

\def\thefootnote{\fnsymbol{footnote}}

\begin{center}
{\large {\bf Medium Baseline Reactor Neutrino Experiments\\ with 2 Identical Detectors
  } }

\bigskip

\bigskip

{\large \noindent Emilio
Ciuffoli$^{1}$\footnote{ciuffoli@ihep.ac.cn}, Jarah
Evslin$^{1}$\footnote{\texttt{jarah@ihep.ac.cn}}, Zhimin Wang$^{2}$\footnote{\texttt{wangzhm@ihep.ac.cn}},\\ Changgen Yang$^{2}$\footnote{\texttt{yangcg@ihep.ac.cn}}, Xinmin Zhang$^{3,
1}$\footnote{\texttt{xmzhang@ihep.ac.cn}} and Weili Zhong$^{2}$\footnote{\texttt{zhongwl@ihep.ac.cn}} }
\end{center}

\renewcommand{\thefootnote}{\arabic{footnote}}

\vskip.7cm

\begin{center}
\vspace{0em} {\em  { 1) TPCSF, IHEP, Chinese Acad. of Sciences\\
2)  Particle astrophysics division, IHEP, Chinese Acad. of Sciences\\
3) Theoretical physics division, IHEP, Chinese Acad. of Sciences\\
YuQuan Lu 19(B), Beijing 100049, China}}

\vskip .4cm

\vskip .4cm

\end{center}


\noindent
\begin{center} {\bf Abstract} \end{center}

\noindent
In the next 10 years medium baseline reactor neutrino experiments will attempt to determine the neutrino mass hierarchy and to precisely measure $\theta_{12}$. Both of these determinations will be more reliable if data from identical detectors at distinct baselines are combined. While interference effects can be eliminated by choosing detector sites orthogonal to the reactor arrays, one of the greatest challenges facing a determination of the mass hierarchy is the detector's unknown energy response.  By comparing peaks at similar energies at two identical detectors at distinct baselines, one eliminates any correlated dependence upon a monotonic energy response. In addition, a second detector leads to new hierarchy-dependent observables, such as the ratio of the locations of the maxima of the Fourier cosine transforms.   Simultaneously, one may determine the hierarchy by comparing the $\chi^{2}$ best fits of $\Delta M^{2}_{32}$ at the two detectors using the spectra associated to both hierarchies. A second detector at a distinct baseline also breaks the degeneracy between $\theta_{12}$ and the background neutrino flux from, for example, distant reactors and increases the effective target mass, which is limited by current designs to about 20 kton/detector.

\vfill

\begin{flushleft}
{\today}
\end{flushleft}
\end{titlepage}

\hfill{}


\setcounter{footnote}{0}


\noindent
\section{Motivation}

10 years ago Petcov and Piai suggested that a medium baseline reactor neutrino experiment can determine the neutrino mass hierarchy \cite{petcovidea} which is manifested as a subtle shift in the locations of peaks in the neutrino spectrum resulting from 1-3 oscillations.  Early studies of this proposal found that it requires an unparalleled precision and an enormous detector \cite{parke2007,caojun2}.  The situation changed with the recent demonstrations \cite{dayabay,neut2012,reno} that 1-3 oscillations are as much as an order of magnitude larger than had been believed just a year earlier. Using the new value of $\theta_{13}$ the analysis of Ref.~\cite{simulazioni} found that the hierarchy could now be determined with a 20 kton detector, which is about the largest that a sufficiently precise detector consisting of two concentric spheres can be.  However it found that, with detectors at the locations proposed in Refs.~\cite{caojunseminario,weihai} neutrinos arriving from reactors at multiple baselines would erase the low energy 1-3 oscillations, diminishing the significance of a determination of the hierarchy.  This problem could be resolved if the detector is placed perpendicular to a reactor array, but at the cost of using the flux from one reactor array instead of two and so increasing the statistical errors.  Another study \cite{oggi} found that a determination of the hierarchy requires a determination of the nonlinear energy response of the detectors to a better precision than has ever been achieved.

Just as Daya Bay and RENO were able to significantly reduce their systematic errors by relying only upon the relative flux observed at distinct baselines, in the present note we will discuss how certain relative measurements with two identical detectors at distinct baselines are insensitive to the detector's correlated  energy response, as has been suggested in Refs.~\cite{simulazioni,DYBIIinternalDoc,teorico,mioweihai}.  For example we will show that the sign of the energy difference between two peaks in the spectra observed at the two detectors can provide a determination of the neutrino mass hierarchy which is independent of the correlated energy responses of the detectors. We also will introduce other 2 additional two-detector observables which are sensitive to the hierarchy and reasonably insensitive to the detector's correlated energy response.

While it may be difficult to build a sufficiently precise single detector with target mass larger than 20 kton, a two detector design can provide 40 kton of target mass. A second detector also yields new observables which can be used to determine the hierarchy.  For example, the ratio of the locations of the global maxima of the Fourier cosine transforms is not only sensitive to the hierarchy, but also reasonably nondegenerate with the observables defined in Refs. \cite{caojun,teorico}.    In addition one may use a $\chi^2$ fit to the spectra with both hierarchies to determine $\Delta M^{2}_{32}$ at both detectors, the correct hierarchy yields the most compatible values for $\Delta M^{2}_{32}$.   Both of these observables are robust in that they are independent of the overall normalization of the detector's energy response.

We will begin in Sec.~\ref{revsez} with a review of reactor neutrino oscillations and how they are affected by the hierarchy.  In Sec.~\ref{nonlinsez} we will explain how a comparison of peak locations at similar energies but distinct baselines may be used to determine the neutrino mass hierarchy with two detectors that have a potentially large unknown correlated nonlinear energy response.  In Sec.~\ref{PeakCounting}, we show that the neutrino mass hierarchy can be determined from the ratio of the oscillation frequencies at the two detectors. In Sec.~\ref{Chisquare2Detector} we describe how a $\chi^2$ fit of $\Delta M^{2}_{32}$ at both detectors provides yet another determination of the hierarchy. In Sec.~\ref{thetasez} we explain that the presence of a second detectors can break the degeneracy between the background electron antineutrino flux and $\theta_{12}$.  Finally in Sec.~\ref{dissez} we will discuss future directions, including the incorporation of these ideas in a more complete simulation of a reactor experiment. 

\section{The electron survival probability} \label{revsez}

Consider for simplicity a single reactor and two detectors at baselines $L$ and $3L/2$.  Optimal values of $L$ will be between 30 and 40 km.
The electron neutrino weak interaction eigenstate $|\nu_e\rangle$ is not an energy eigenstate $|k\rangle$, but it can be decomposed into a real sum of energy eigenstates
\beq
|\nu_e\rangle=\c12\c13|1\rangle+\s12\c13|2\rangle+\s13|3\rangle.
\eeq
In the relativistic limit, after traveling a distance $L$, the survival probability of a coherent electron (anti)neutrino wavepacket with energy $E$ can be expressed in terms of the mass matrix $\mathbf{M}$
\bea
P_{ee}&=&|\langle\nu_e|\mathrm{exp}\left(i\frac{\mathbf{M}^2L}{2E}\right)|\nu_e\rangle|^2\label{pee}\\
&=&\sp413+\cp412\cp413+\sp412\cp413+\frac{1}{2}(P_{12}+P_{13}+P_ {23})\nonumber\\
P_{12}&=&\spp2212\cp413\cos\left(\frac{\m21L}{2E}\right)\hsp
P_{13}=\cp212\spp2213\cos\left(\frac{\m31L}{2E}\right)\nonumber\\
P_{23}&=&\sp212\spp2213\cos\left(\frac{\m32L}{2E}\right)\nonumber
\eea
where $\m{i}{j}$ is the mass squared difference of mass eigenstates $i$ and $j$.

\begin{figure} 
\begin{center}
\includegraphics[width=5.5in,height=2.2in]{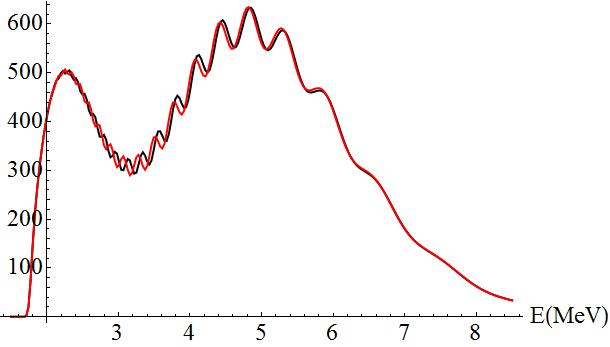}
\caption{Theoretical spectra of 6 years live time of neutrinos at 20 kton detectors a baseline of 54 km from a 23.2 GW reactor.   The vertical axis is the number of neutrinos expected in each 30 keV bin.  A detector resolution of $2.5\%/\sqrt{E/{\rm{MeV}}}$ is assumed.  The same value of $\meff$, defined in Eq.~(\ref{meff}), is used plotting both the normal (red) and inverted (black) hierarchies, which is consistent with the MINOS measurements in Ref.~\cite{minosneut2012}.  This fixes the high energy parts of the two spectra to be nearly identical.}
\label{teorfig}
\end{center}
\end{figure}

\section{Comparing peak locations} \label{nonlinsez}

\subsection{The peak location}

The survival probability at large distances is dominated by 1-2 oscillations, described by $P_{12}$ with a fine structure of smaller $P_{13}$ oscillations which are slightly perturbed by the yet smaller $P_{23}$ oscillations.  This fine structure is used to determine the hierarchy as can be seen in Fig.~\ref{teorfig}.  If $P_{23}$ were constant, then the fine structure peaks would be determined entirely by $P_{13}$ and so would be periodic in $L/E$-space.  The $n$th peak in the neutrino spectrum\footnote{The zeroth peak corresponds to infinite energy, at which neutrinos have not oscillated.  Higher $n$ corresponds to lower neutrino energy.  The first few peaks are invisible at medium baselines due to the low reactor flux at the corresponding high energies.}, corresponding to neutrinos that have oscillated $n$ times, would be located at the energy
\beq
E^{(0)}_n(L)=\frac{\mn31 L}{4\pi n}.
\eeq
However the $P_{23}$ oscillations deform this periodicity, shifting the $n$th peak to the energy
\beq
E_n(L)=\frac{\mn31 L}{4\pi(n\pm\an)} \label{en}
\eeq
where the plus (minus) sign corresponds to the normal (inverted) hierarchy and the perturbations $\an$ can be determined using Eq.~(\ref{pee}).  The $\an$\ depend weakly upon unknown neutrino mixing parameters, but as a determination of the hierarchy is equivalent to a determination of the sign with which $\an$ enters (\ref{en}), this small uncertainty in its value will be irrelevant.

\subsection{Comparing the 10th and 15th peaks}

It follows from Eq. (\ref{en}) that the ratio of the energy of the 15th peak at $3 L/2$ to that of
the 10th peak at $L$ is
\beq
\frac{E_{15}(3L/2)}{E_{10}(L)}=\frac{3(10\pm\alpha_{10})}{2(15\pm\alpha_{15})}\sim 1\pm
\frac{3\alpha_{10}-2\alpha_{15}}{30}. \label{fraz}
\eeq
For ease of comparison with previous studies we will use old values of the neutrino mass matrix parameters 
\beq
\m21=7.59\times 10^{-5}{\mathrm{\ eV^2}}\hsppp
\mn32=2.4\times 10^{-3}{\mathrm{\ eV^2}}\hsppp
\spp2212=0.8675\hsppp
\spp2213=0.092.
\eeq
Then the relevant $\an$ can be read from Fig. 1 of Ref.~\cite{teorico}
\beq
\alpha_{10}=0.072\hsp \alpha_{15}=0.035
\eeq
which can be substituted into Eq. (\ref{fraz}) to yield the energy ratio
\beq
\frac{E_{15}(3L/2)}{E_{10}(L)}=1.000\pm 0.005
\eeq
where again the positive (negative) sign corresponds to the normal (inverted) hierarchy.

Therefore we have learned that in the case of the normal (inverted) hierarchy the energy of the 15th peak at the far detector will be 0.5\% higher (lower) than that of the 10th peak at the near detector.  A 0.5\% difference is small and difficult to measure with limited statistics and detector response.  By the 17th peak this difference increases to 1\%, but the detector resolution is reduced at these energies.  However the important point is that since the energy response of the detector is monotonic with respect to the true energy, the observed energy of the far detector peak will be greater (less) than the observed energy at the near detector if and only if the true energy is greater (less) which indicates a normal (inverted) hierarchy.  Thus this determination of the hierarchy is independent of the unknown correlated nonlinear energy response of the detector.

\subsection{How to find the 10th and 15th peaks}

There is however one complication.  In order to use this technique, one needs to be able to identify the 15th peak at the far detector and the 10th peak at the near detector.  How can this be done?  The positions of these peaks depend on $\mn32$ whose measurement at MINOS has error bars only half as large as the distance between the 15th and 16th peaks.

To find these peaks, we suggest that one begin with the 6th peak at the near detector.  This is easy to find, using Eq. (\ref{en}) it lies at an energy of
\beq
E_6(L)=\frac{\mn31 L}{4\pi(6\pm\alpha_6)}\sim\frac{\mn31 L}{4\pi(6\pm 0.052)}.
\eeq
The relative energy difference between the 6th and 7th peak is 17\%, about four times larger than the error with which $\mn32$ is known from MINOS.  To better estimate the errors, one can use the asymptotic form of $\an$ at low values of $n$~\cite{parke2005}
\beq
E_6(L)=\frac{\meff L}{24\pi}.
\eeq
where
\beq
\Delta M^2_{eff}=\cp212\mn31+\sp212\mn32. \label{meff}
\eeq
The combination $\meff$ is determined by MINOS with an error of about 5\%.  Thus the 6th peak
at the near detector can be distinguished from the 7th peak at the $3\sigma$ level.   Including a nonlinear response of 2 to 3 percent the reliability of this determination is reduced to the $2\sigma$ level.

\begin{figure} 
\begin{center}
\includegraphics[width=5.5in,height=2.2in]{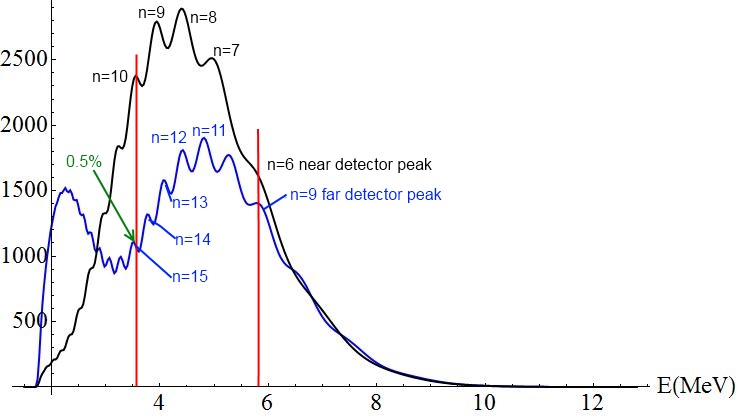}
\caption{Theoretical spectra of 6 years of neutrinos at 20 kton detectors at baselines 36 km (black) and 54 km (blue) from a 23.2 GW reactor.   The vertical axis is the number of neutrinos expected in each 30 keV bin, although the far detector flux has been tripled to render its features visible in this figure.  A detector resolution of $2.5\%/\sqrt{E/{\rm{MeV}}}$ is assumed.  The 6th peak at the near detector lies at virtually the same energy as the 9th peak at the far detector.  However the 15th far peak is at a slightly lower energy than the 10th near peak, indicating an inverted hierarchy.}
\label{indfig}
\end{center}
\end{figure}

Fortunately, the determination can be made more precise.  For one, NO$\nu$A is likely to measure $\mn32$ more precisely before Daya Bay II produces results.  More importantly, while the 6th peak at the near detector corresponds to the 9th peak at the far detector\footnote{The hierarchy makes little difference at such low values of $n$ because $\alpha_n$ is very close to linear, indeed $\alpha_9/\alpha_6$ is 1.42.}, the 5th and 7th peaks both correspond to minima.  Thus a misidentification of the 6th peak at the near detector can be discovered if, at the same energy, there is a minimum in the far detector flux.  Again these relative comparisons between the near and far detector flux are independent of the unknown nonlinear response to the extent that the near and far detectors are identical.

Thus it seems possible to identify the sixth peak of the near detector with about $4\sigma$ of certainty.  To find the tenth peak, one then needs to count peaks.  One knows the true energy difference between peaks from Eq. (\ref{en}), although this can be somewhat distorted by nonlinearities.  While counting peaks at the far detector may be difficult in practice due to the low flux, counting peaks at the near detector at baselines of less than 40 km is feasible, although perhaps less reliable than a method which uses all of the intermediate information like a $\chi^2$ fit.

\subsection{The Procedure}

Summarizing, the determination of the hierarchy is a four step process:

\noindent
{\bf Step 1:} Use Eq. (\ref{en}) to identify the 6th peak at the near detector.

\noindent
{\bf Step 2:} Check that at the same energy one finds a peak at the far detector, if not, the sixth peak at the near detector is the second nearest peak to the result of  Eq. (\ref{en}).

\noindent
{\bf Step 3:} Using the expected distance between peaks from Eq. (\ref{en}) and the location of the sixth peak found above, count peaks to find the tenth peak at the near detector.

\noindent
{\bf Step 4:} The measured energy of the 15th peak at the far detector will either be 0.5\% higher or lower than the measured energy of the 10th peak at the near detector.  If it is higher (lower) than the neutrino mass hierarchy is normal (inverted).

\begin{figure} 
\begin{center}
\includegraphics[width=5.5in,height=2.1in]{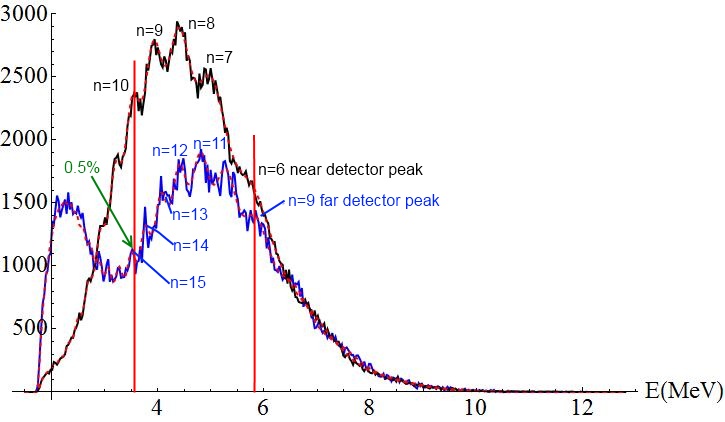}
\caption{Typical simulated spectra of 6 years of neutrinos at 20 kton detectors at baselines 36 km (black) and 54 km (blue) from a 23.2 GW reactor.   The red curves are the theoretical spectra of Fig.~\ref{indfig}.  The vertical axis is the number of neutrinos expected in each 30 keV bin, although the far detector flux normalization has been rescaled by a factor of 3 in the figure to render its features visible.  A detector resolution of $2.5\%/\sqrt{E/{\rm{MeV}}}$ is assumed. While the peaks discussed in this paper are visible, their positions are difficult to determine by eye to within the required 0.5\%. }
\label{simfig}
\end{center}
\end{figure}

These steps are illustrated on the theoretical spectra, which exclude statistical errors, in Fig.~\ref{indfig}.  This figure contains the expected spectra at 36 and 54 km in the case of the inverted hierarchy assuming an optimistic energy resolution.  The nonlinear detector response is not included in this figure.  To see the effects of statistical errors, we have included the results of a simulation with the same parameters in Fig.~\ref{simfig}.  Here we have considered 6 years live time of neutrinos arising from a reactor complex with a thermal capacity of 23.2 GW, corresponding roughly to the pairs of reactors at Daya Bay, Ling Ao I and Ling Ao II plus the planned pair Ling Ao III.  Each detector is assumed to be 20 ktons.  While the construction of a Daya Bay II like detector, made from concentric spherical shells, larger than 20 ktons seems difficult if not impossible, the construction of 2 detectors allows an effective target mass of 40 ktons and so, in addition to helping with the nonlinearity problem, also decreases the fractional statistical errors in the neutrino flux.  We have simulated detector locations perpendicular to the reactor array\footnote{In the case of Daya Bay this would correspond roughly to HuaShan and GuanYin mountain park respectively.}, so that the baselines to various reactors are essentially identical, thus avoiding the interference effects of Ref.~\cite{teorico}.

\section{Comparing oscillation frequencies} \label{PeakCounting}

We have seen that a comparison of individual peak energies can be independent of the correlated detector nonlinear energy response.  In practice, given the long baselines which are necessary, a determination of the hierarchy will be strongly limited by statistics.  Therefore the best determination of the hierarchy uses information from the entire spectrum, not just from a few peaks.  While the determination described above can and should be applied to every visible peak, the highest sensitivity to the hierarchy arises at low energies where the individual peaks are difficult to identify.

There are two known ways to take advantage of the information stored in the entire spectrum, including the invisible peaks.  One can use a Fourier transform to sum them together but a different direction to Ref.~\cite{hawaii} or one may perform a $\chi^2$ analysis.  In the next two sections we will describe new, hierarchy-dependent observables which are provided by a second detector using these two methods respectively.

\subsection{The oscillation frequencies}

\begin{figure} 
\begin{center}
\includegraphics[width=4in,height=2in]{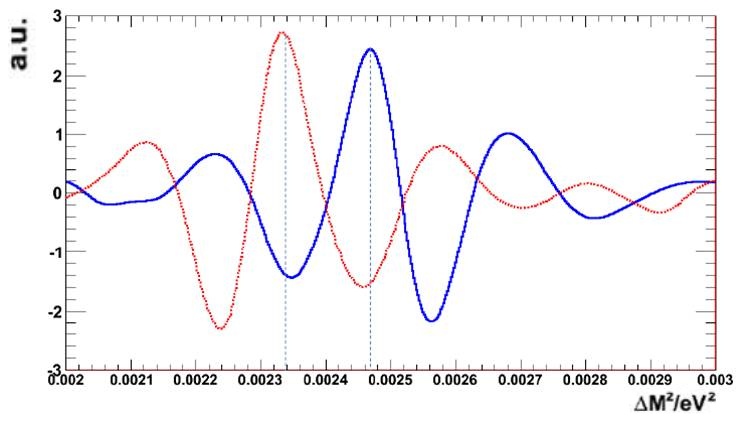}
\caption{The theoretical cosine Fourier transform spectrum of the oscillated reactor neutrino spectrum at 60km with a $3\%/\sqrt{E/{\rm{MeV}}}$ visible energy resolution, blue curve: normal hierarchy; red curve: inverted hierarchy.}
\label{FourierExampleFCT}
\end{center}
\end{figure}

The Fourier cosine transform of the observed neutrino spectrum $\Phi(L/E)$ at energy $E$ at the $i$th detector,  depicted in Fig.~\ref{FourierExampleFCT} in the case of a $L=60$\ km baseline, is
\beq
F(k)=\int d\left(\frac{L}{E}\right) \Phi(L/E)  \cos\left(\frac{kL}{2E}\right).
\eeq
This transform exhibits a global maximum at a frequency $\Delta M_i^{2}$ near $\mn31$ \cite{caojun}.  The location of this maximum  can be determined using the results of Ref.~\cite{teorico}.  

At baselines of 30 km or less,  neutrinos of high enough energy to be observed using inverse $\beta$ decay have oscillated less than 10 times.  These first oscillations occur with a frequency determined entirely by $\meff$ \cite{parke2005} and not by the hierarchy.   The maximum of the Fourier cosine transform of the near detector spectrum lies at just this frequency
\beq
\Delta M_{\rm{near}}^{2}=\meff=\cp212\mn31+\sp212\mn32.
\eeq

On the other hand, at a baseline of 60 km one can in principle detect at least the first 16 oscillations.  The average $L/E$ frequency of the first 16 oscillations is $\mn31$ therefore at a far detector the maximum of the Fourier cosine transform lies at approximately
\beq
\Delta M_{\rm{far}}^{2}\sim\mn31.
\eeq
The precise location of the maximum depends on just how many peaks can be discerned at the detector.  At low energies the detector's energy resolution leads to a decrease in the peak amplitudes, and the result is that the low energy peaks contribute less to the nonzero frequency part of the Fourier transform.  The low energy peaks of the spectrum drive the maximum of the Fourier transformed spectrum from $\meff$ towards and possibly beyond $\mn31$.   Thus the better the energy resolution, the further $\Delta M_{\rm{far}}^{2}$ will be from $\meff$.

The hierarchy is determined by the ratio
\beq
f_{\rm{dis}}=\frac{\Delta M_{\rm{near}}^{2}}{\Delta M_{\rm{far}}^{2}}\sim\frac{\meff}{\mn31}=1+\sp212\left(\frac{\mn31-\mn32}{\mn31}\right). \label{DisPeakCount}
\eeq
Roughly speaking, a value greater than $1$ indicates the normal hierarchy and less than one indicates the inverted hierarchy.  In practice the resolution determines an overall shift in $\Delta M_{\rm{far}}^{2}$ and so the threshold value will not be precisely equal to $1$, but can be determined for example via simulations.



\subsection{Results}

\begin{figure} 
\begin{center}
\includegraphics[width=4in,height=2in]{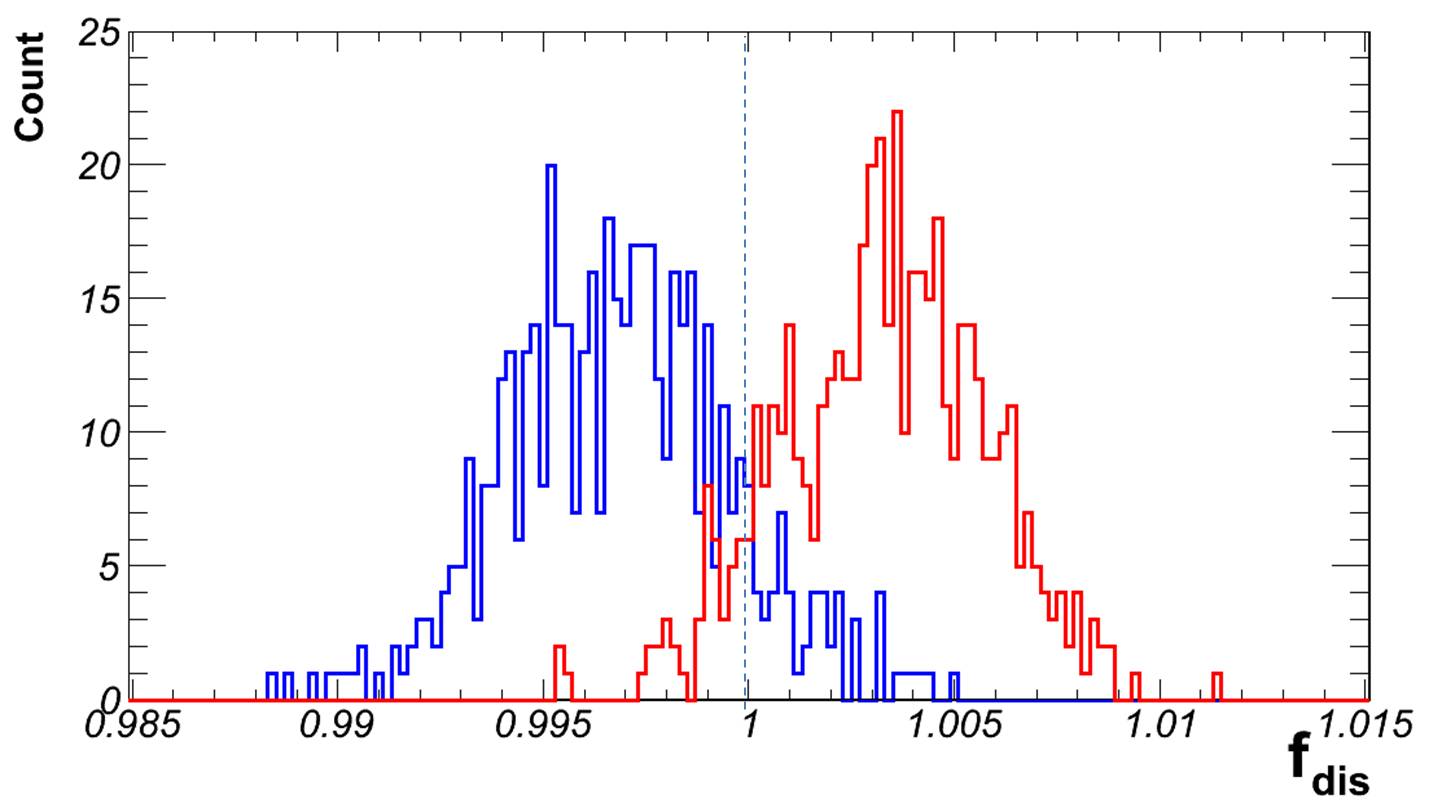}
\caption{Histogram of the number of simulated experiments yielding each value of $f_{\rm{dis}}$ for a near baseline of $L_{\rm{near}}=30$ km and a far baseline of $L_{\rm{far}}=60$ km, Blue curve: normal hierarchy; Red curve: inverted hierarchy. The dashed line is the threshold which yields the hierarchy determination.}
\label{DistributionPeak30km}
\end{center}
\end{figure}

As an illustration of this method, consider the neutrinos from a set of reactors with a total thermal capacity of $\sim$32 GW all of which are 30 km from a 20 kton near detector and 60 km from an identical far detector.  We have simulated 500 experiments assuming each hierarchy, with a detector visible energy resolution of $3\%/\sqrt{E/{\rm{MeV}}}$.  We have assumed 50k events at the far detector and 200k at the near site.   This corresponds to roughly $\sim$3 years live time\footnote{The large backgrounds expected at such an experiment \cite{caojunshenzhen} may require severe cuts and thus an imperfect detector efficiency, as a result 3 years of live time may require appreciably more than 3 years of running at sites with only 500 meters of rock overburden.}, neglecting the decrease in total flux due to 1-2 oscillations.   We have not considered the unknown nonlinear detector response and errors here.

The distribution of $f_{\rm{dis}}$ is shown in Fig.~\ref{DistributionPeak30km}.  The dashed line is the identification cut, which appears to be very close to 1. The resulting probabilities of success are summarized in Table.~\ref{DisPeakRatio}, where they are compared with the peak-valley (RL+PV) analysis of Ref.~\cite{caojun2} using one or two 20 kton detectors at 60 km.  While there is no reason to expect a perfect symmetry, the large asymmetry between the chance of success in the case of the two hierarchies may be an artifact of the samples of experiments considered\footnote{In Table.~\ref{DisPeakRatio}, we used the RL+PV method with a threshold values of $-0.014$ and $-0.04$ respectively for 1 or 2 detectors at 60km (the first and fourth rows).  These thresholds have been chosen to yield  the maximum average probabilities of successfully determining the hierarchy in a sample of experiments with an equal number of NH and IH cases.   In the second row of the table, the threshold used for $f_{\rm{dis}}$  is 1, and in the third row the threshold for $RL+PV$ is $37.5\times f_{\rm{dis}}-37.55$.}.

\begin{table}[position specifier]
\centering
\begin{tabular}{|c|l|l|}
\hline
Method&NH(\%)&IH(\%)\\
\hline\hline
RL+PV&&\\
20kton$\times$3years&85.2&89.7\\
\hline
$f_{\rm{dis}}$&&\\
Oscillation frequency&89.2&90.0\\
\hline
RL+PV\&$f_{\rm{dis}}$&&\\
Combination&96.9&95.3\\
\hline
RL+PV&&\\
20kton$\times$2$\times$3years&93.2&93.2\\
\hline
\end{tabular}
\caption{The probability of successfully determining hierarchy with 20 kton/detector and 3 years live time using the Fourier peak-valley analysis (RL+PV) at a 60 km baseline detector, the oscillation frequency method ($f_{\rm{dis}}$) using a 30 km and a 60 km detector, a combination of both (RL+PV\&$f_{\rm{dis}}$) and finally only RL+PV but with 2 detectors at 60 km.}
\label{DisPeakRatio}
\end{table}

\begin{figure} 
\begin{center}
\includegraphics[width=4in,height=2in]{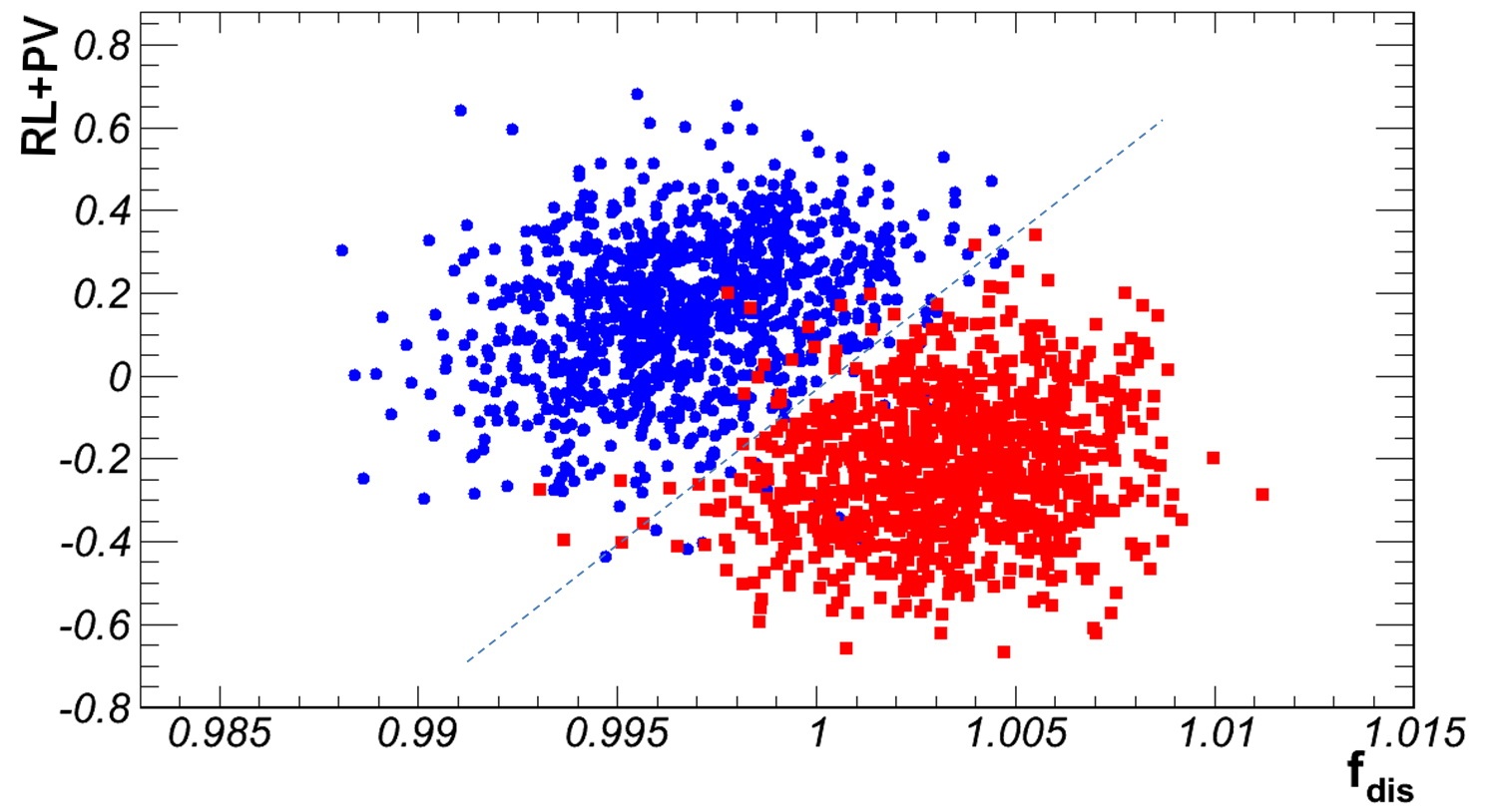}
\caption{The correlation between the Fourier Transform peak-valley analysis (RL+PV method) and the oscillation frequency analysis ($f_{\rm{dis}}$).}
\label{Distribution2DPeakVsPv}
\end{center}
\end{figure}

In the third row of Table.~\ref{DisPeakRatio} we consider a combination of the 20 kton 60 km  RL+PV analysis and the oscillation frequency analysis presented here.  This combination, which uses a 20 kton detector at 30 km and at 60 km, significantly outperforms the two 20 kton detectors at 60 km analyzed in the fourth row.  The strong improvement attained by combining the RL+PV and oscillation frequency methods is a result of the fact that they are only weakly degenerate, as can be seen in Fig. \ref{Distribution2DPeakVsPv}.


As our method relies only upon the {\it{relative}} positions of the peaks in the Fourier transformed spectra, it is not affected by the uncertain value of $\mn32$ \cite{hawaii}.  Furthermore, it was shown in Ref.~\cite{petcov2010} that the Fourier analysis is not affected by a rescaling of the energy and hardly affected by a constant energy shift, and therefore is robust with respect to a non-linear energy response of the form $E_{rec}=a\times E+b$ and even reasonably independent of the reactor flux model. We also have confirmed that the oscillation frequency method is more sensitive to statistics especially at the far site, for example if we have more than 100k events at far site then the oscillation frequency method outperforms the Fourier peak-valley analysis.

\section{Comparing fitted $\Delta M^{2}_{32}$} \label{Chisquare2Detector}

As has been shown in Ref.~\cite{qianstat}, if interpreted correctly, a $\chi^2$ analysis can be used to determine the neutrino mass hierarchy at a reactor experiment.  As it uses all of the information available in the spectrum, it can potentially provide a more robust determination of the hierarchy than a Fourier analysis, as was seen for example in the simulations of Ref.~\cite{oggi}.

The $\chi^2$ statistic corresponding to the simulated spectrum $N_{{\rm{spectrum}},i}$ is defined to be
\beq
\chi^{2}=\sum_{i}^{n}\frac{(N_{{\rm{spectrum}},i}-N_{{\rm{fitted}},i})^{2}}{N_{{\rm{spectrum}},i}} \label{Chi2Definition}
\eeq
where $N_{{\rm{fitted}},i}$ is the best fit spectrum with a given hierarchy.  For simplicity we will only fit the parameter $\mn32$.  Here the index $i$ labels the 50 keV bins.  The visible energy resolution is taken to be $3\%/\sqrt{E/{\rm{MeV}}}$.



\begin{figure} 
\begin{center}
\includegraphics[width=3.1in,height=1.1in]{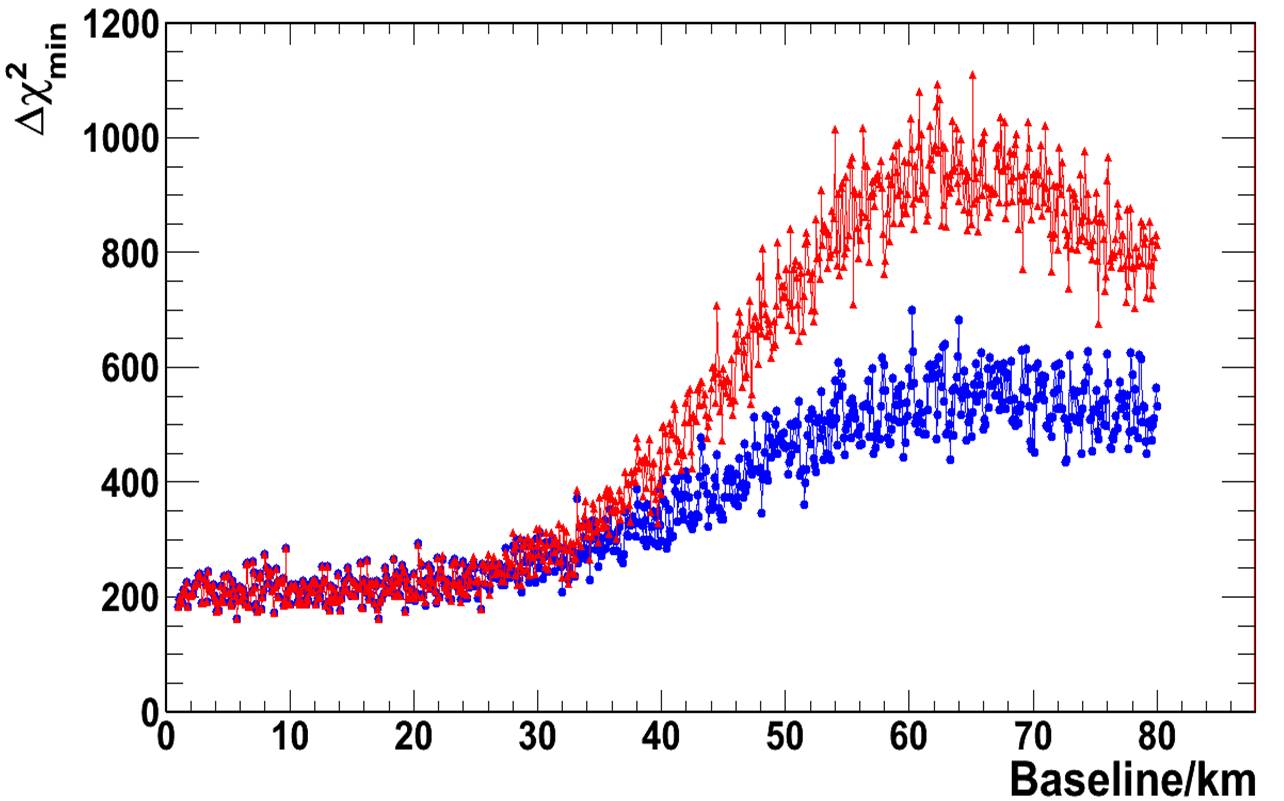}
\includegraphics[width=3.1in,height=1.1in]{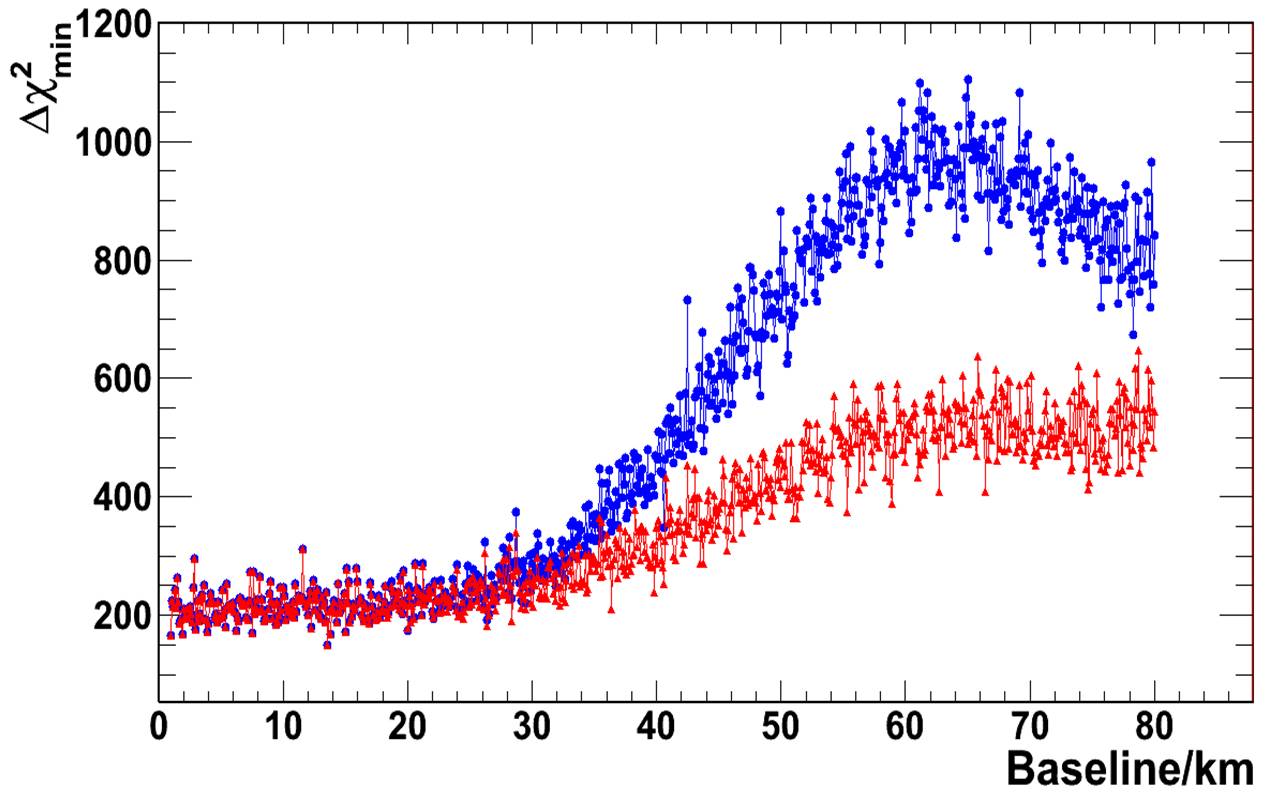}
\includegraphics[width=3.1in,height=1.1in]{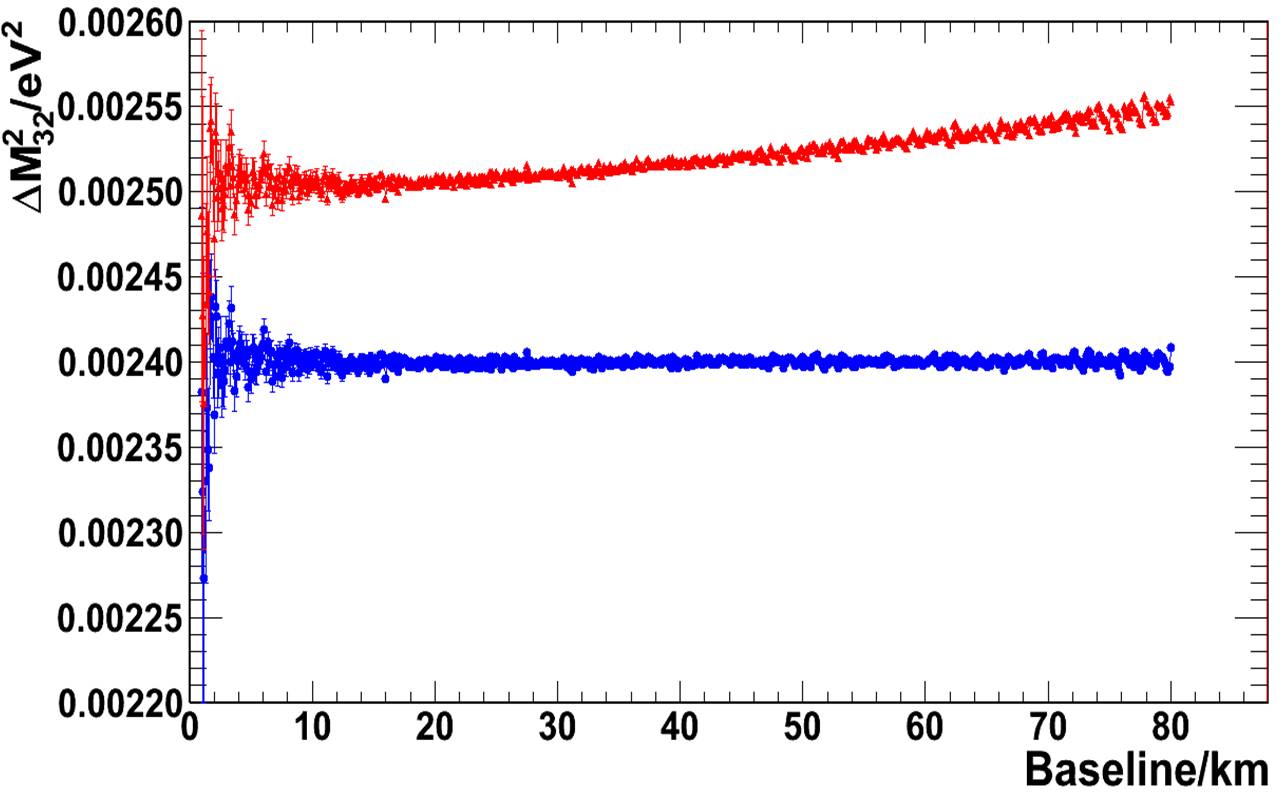}
\includegraphics[width=3.1in,height=1.1in]{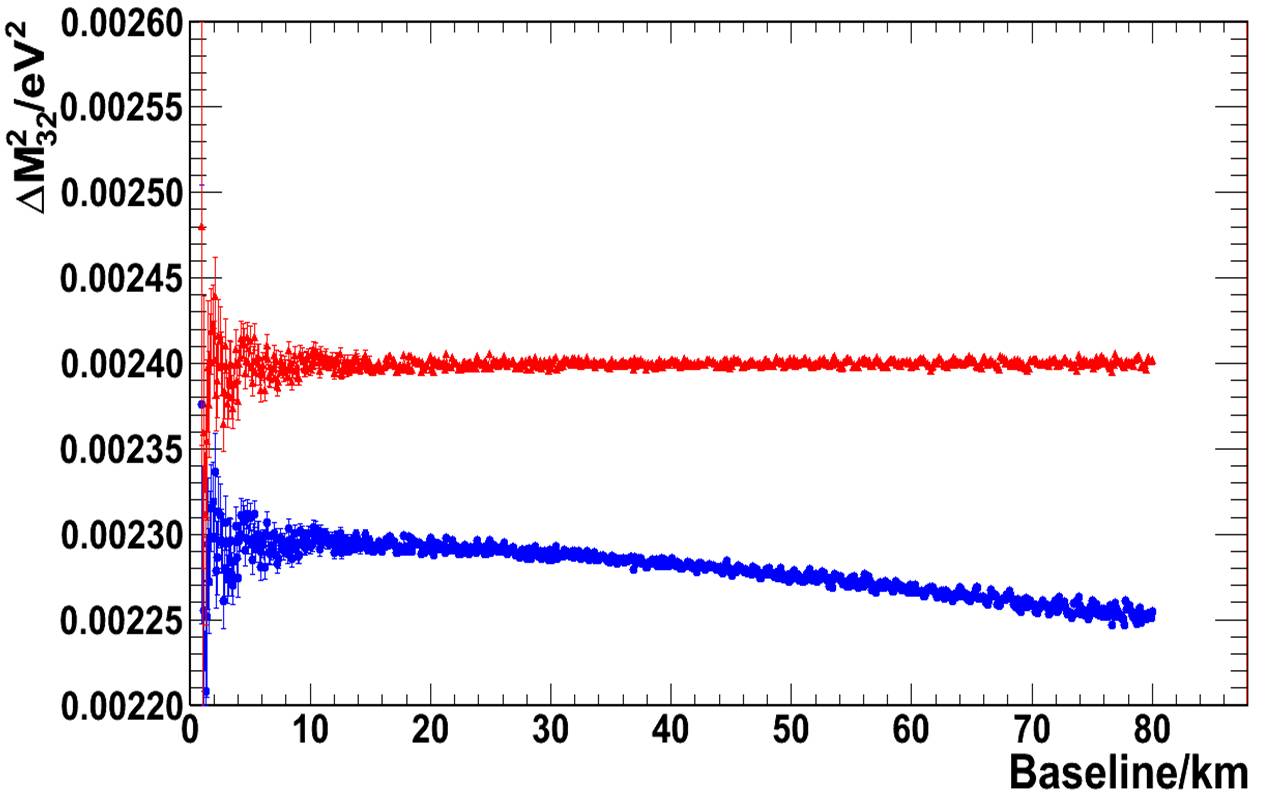}
\caption{The bottom panels are the fit values of $\Delta M^{2}_{32}$ at various baselines assuming one million events.  The top panels are the corresponding values of $\chi^2_{min}$. left: simulated data using the normal hierarchy, blue: fit with NH; Red: fit with IH; right:  simulated data using the inverted hierarchy, blue: fit with NH; Red: fit with IH.}
\label{ChiVsBaseline}
\end{center}
\end{figure}

We have found the value of $\mn32$ which minimizes the $\chi^2$ value of fits to both hierarchies at various baselines.  As can be seen in Fig.~\ref{ChiVsBaseline}, while a fit to the correct hierarchy, by construction, correctly yields the simulated value $\mn32=2.4\times 10^{-3}{\mathrm{\ eV^2}}$, a fit of the NH (IH) simulated data to the theoretical IH (NH) spectrum yields a best fit for $\mn32$ which depends upon the baseline.   This strong baseline dependence of the best fit persists even at short baselines, where fits to both hierarchies yield comparable values of $\chi^2$.


This observation may be used to determine the hierarchy as follows.  One may use a $\chi^2$ fit of the spectra observed at two different baselines, assuming both hierarchies, to generate 4 values of $\mn32$.  The values generated by the correct hierarchy at the various detectors will all agree, whereas those generated by the wrong hierarchy will depend upon the baseline.  For example, in the case simulated here the values of $\mn32$ resulting from these fits are summarized in Table~\ref{DeltaM32VsBaseline}.  Here we have also included the corresponding projected fits from the Daya Bay experiment's far detectors with simulated data, which similarly yield a hierarchy dependent fit at an effective baseline.  The hierarchy is determined from the relative values of $\mn32$, which in the case of identical detectors is less sensitive to the correlated nonlinear energy response.



\begin{table}[position specifier]
\centering
\begin{tabular}{|c|l|l|l|l|l|}
\hline
 & &\multicolumn{2}{|c|}{NH $\Delta M_{32}^{2}\pm\varepsilon(10^{-3}eV^{2})$}&\multicolumn{2}{|c|}{IH $\Delta M_{32}^{2}\pm\varepsilon(10^{-3}eV^{2})$}\\
\hline
Baseline&Statistics&fit by NH&fit by IH&fit by NH&fit by IH\\
\hline\hline
1.5km&50k&2.440$\pm$0.160&2.540$\pm$0.160&2.185$\pm$0.190&2.290$\pm$0.170\\
\hline
30km&50k&2.417$\pm$0.008&2.521$\pm$0.010&2.291$\pm$0.006&2.402$\pm$0.009\\
\hline
60km&50k&2.394$\pm$0.003&2.528$\pm$0.004&2.270$\pm$0.004&2.404$\pm$0.004\\
\hline
1.5km&1M&2.349$\pm$0.036&2.452$\pm$0.038&2.294$\pm$0.035&2.398$\pm$0.037\\
\hline
30km&1M&2.398$\pm$0.002&2.509$\pm$0.002&2.288$\pm$0.002&2.399$\pm$0.001\\
\hline
60km&1M&2.400$\pm$0.001&2.533$\pm$0.001&2.268$\pm$0.001&2.398$\pm$0.001\\
\hline
\end{tabular}
\caption{Best fit $\Delta M_{32}^{2}$ value versus baseline for NH and IH, which all are based on the simulated data with 1M events and $3\%/\sqrt{E/{\rm{MeV}}}$ visible energy resolution, without consideration of backgrounds and errors.}
\label{DeltaM32VsBaseline}
\end{table}


\section{Background Flux and $\theta_{12}$} \label{thetasez}

The existence of a second detector at a distinct baseline can also help with another of Daya Bay II's goals \cite{caojunseminario}, the determination of the mixing angle $\theta_{12}$. This is determined by ignoring the small 1-3 oscillations in the observed oscillated reactor neutrino spectrum, and measuring the depth of the flux minimum at the energy corresponding to the 1-2 oscillation maximum. A precise determination of $\theta_{12}$ can provide a powerful test of the unitarity of the neutrino mass matrix, providing an indirect probe of various sterile neutrino models \cite{xing}.

Proposed sites for the construction of medium baseline reactor experiments are in China's Guangdong province and in South Korea, both of which are experiencing the rapid construction of nuclear reactors.  Only the nearest reactors can effectively be used to determine the hierarchy, more distant reactors provide backgrounds.  The Daya Bay II location of Ref.~\cite{caojunseminario} may have a large background from the proposed reactor at Lufeng, and the perpendicular locations proposed in Ref.~\cite{simulazioni} may have backgrounds from the proposed reactor at Huidong.   As described in Ref.~\cite{teorico}, while the background reactors are twice as far as the foreground reactors, they may actually provide more flux at the 1-2 minimum than the foreground reactors because at that energy the background reactors are at their 1-2 minimum.  In addition, the other sites proposed for Daya Bay II, in western Guangdong, along with all possible sites for Reno 50 \cite{reno50} will have to contend with background reactors 200 km away, whose contribution to the flux at the 1-2 maximum is nonetheless appreciable \cite{mioweihai}.

Can this background flux be simply subtracted away?  Of course its contribution to the statistical error cannot be removed \cite{petcov87}.  But also it contributes a systematic error, as
the overall reactor flux normalization is poorly understood \cite{reattoreanom} and so the expected background flux cannot be reliably determined. With the vast array of experiments underway it will no doubt be measured more precisely before a medium baseline experiment is built, but nonetheless the overall reactor flux normalization is unlikely to be measured as precisely as the 0.63\% desired precision for the measurement of $\spp2212$ at Daya Bay II \cite{caojunshenzhen}.


If the reactor flux normalization is increased, then the background flux at the 1-2 maximum will be increased and so the measured value of $\theta_{12}$ will decrease.  This leads to a degeneracy between $\theta_{12}$ and the overall reactor flux normalization.  This degeneracy will be broken if there are multiple detectors at distinct baselines, because the relative flux deficit at the 1-2 maximum measured by each detector will be a distinct combination of the reactor flux normalization and the disappearance due to 1-2 oscillations.  Thus systematic errors due to the unknown reactor flux normalization will be decreased.  For example, the detectors described above at 36 km and 54 km perpendicular to the Daya Bay complex would satisfy both of these criteria.

\section{Discussion} \label{dissez}

In this note we have introduced several multidetector observables which are sensitive to the hierarchy.  The definitions chosen have not yet been optimized.  Thus while the chance of success based on these techniques is overestimated by the fact that various backgrounds and systematics have not been included in our analysis, it is also underestimated as the analyses considered can easily be improved.

For example, consider the oscillation frequency method introduced in Sec. \ref{PeakCounting}.  While the fine structure oscillation frequency observed at the near detector is everywhere $\meff$, that observed at the far detector is energy dependent.  It will be $\meff$ at high energies, and it will deviate from $\meff$ by as much as $3\%$ near the 1-2 oscillation maximum.  The Fourier transform maximum that we have defined in Eq. (\ref{DisPeakCount}) is only sensitive to the average frequency, and so is only sensitive to the hierarchy at the 1\% level, as can be seen in Fig. \ref{DistributionPeak30km}.  Therefore a higher sensitivity may be obtained by Fourier transforming only the part of the spectrum which is the most sensitive to the hierarchy, the low-mid energies corresponding to the 1-2 oscillation maximum.  Such a restricted Fourier transform may be obtained by using a {\it{weighted}} Fourier transform of the kind introduced in Ref.~\cite{nuisance} and used in Ref.~\cite{simulazioni}.   In Ref.~\cite{nuisance} it was shown that such weights are anyway necessary to remove a spurious dependence upon the high energy tail of the neutrino spectrum, which is poorly understood, irrelevant to the hierarchy and a factor of 2-3 times lower than that given by the quadratic fit spectrum \cite{vogelengel} used in studies such as this one and Refs.~\cite{caojun,caojun2}.

\noindent
\textbf{Future directions}

The simulations of Ref.~\cite{simulazioni} show that, even if the detector response is perfectly understood, the chance of determining the hierarchy in 6 years with a single 20 kton detector in Guangdong province appears to be limited to about 96\% using a Fourier analysis, although this can be somewhat improved using a $\chi^2$ analysis \cite{oggi}.  However in this note we have restricted our attention to possible solutions to the nonlinearity problem, considering an idealized situation in which the baselines can be chosen at will, there are no backgrounds and the neutrino flux arises from a single reactor, thus avoiding multiple baseline interference effects.  In this context we can in principle determine whether the seemingly unachievable experimental requirements that Ref.~\cite{oggi} found for the understanding of the nonlinear response at a single detector also apply to multiple detectors, or if instead a multidetector setup can determine the hierarchy in the presence of an unknown nonlinear response larger than that allowed in a single detector setup.

In practice statistical errors are a serious problem, and so an optimal $\chi^2$ fit should introduce pull parameters for the various sources of nonlinearity together with different penalty terms for correlated and uncorrelated errors.  The uncorrelated errors are expected to be subdominant.  The correlated errors on the other hand, as we have argued in this note, are not necessarily a serious obstruction to the determination of the hierarchy at a multidetector experiment.

\section* {Acknowledgement}

\noindent
The authors are honored to express their gratitude to S\"oren Jetter, Liang Zhan and especially Xin Qian for useful discussions and explanations.
JE is supported by the Chinese Academy of Sciences
Fellowship for Young International Scientists grant number
2010Y2JA01. EC and XZ are supported in part by the NSF of
China.


\end{document}

 that $\theta_{13}$ might be large in 2011, in previous years the tight constraints placed by CHOOZ~\cite{chooz} and Palo Verde~\cite{paloverde} meant that analyses of 1-3 oscillations often assumed a value of the mixing parameter $\spp2213$ an order of magnitude or more below the true value.  The new, large value of $\theta_{13}$ has led to a reexamination of the longstanding belief that, with the known low value of $\m21$, it is not practical to measure the neutrino hierarchy at a medium baseline reactor experiment~\cite{petcovidea,parke2007}.  Now not only is such a measurement in principle possible, but indeed it will be attempted within the next decade~\cite{caojunseminario,renonuturn,yifangseminario}.

Using the old value of $\theta_{13}$, the authors of Ref.~\cite{caojun} proposed an optimal baseline and analysis for determining the hierarchy, building upon the Fourier transform strategy of Ref.~\cite{hawaii}.   Then in Ref.~\cite{caojun2} they performed a series of simulations to test these ideas.  In Ref.~\cite{noi} we have updated the analysis of Ref.~\cite{caojun} to the new value of $\theta_{13}$, including several important new effects such as the interference caused by the spacing between reactors in a reactor complex.  In the current note we will update the simulations of Ref.~\cite{caojun2} to the new value of $\theta_{13}$.  We will apply the interference effect and also the spurious dependence of the Fourier transform on high energy neutrinos~\cite{oggi,noispurioso} to study the optimal baseline for such an experiment as well as the optimal mountain under which such an experiment might be built in China's Guangdong province.

As the Daya Bay and Ling Ao reactor complexes enjoy 8 neutrino detectors, they are optimal neutrino sources for a medium baseline reactor experiment.  However the distance between the Daya Bay reactor complex and the two Ling Ao complexes is more than a kilometer, which means that low energy reactor neutrinos can oscillate half of a phase while traveling from one to the other.  As a result, if a detector is roughly along the line which separates these detectors, then the low energy 1-3 neutrino oscillation amplitude will decrease appreciably and so be difficult to observe.  We will refer to this effect as interference, although the neutrinos from Daya Bay and Ling Ao are never coherent, the effect is caused by the addition of probabilities and not wavefunctions.  Since the determination of the hierarchy at a medium baseline experiment~\cite{petcovidea} relies upon the observation of 1-3 oscillation peaks, and in particular the low energy peaks are necessary to break a degeneracy present in the high energy peaks~\cite{parke2007,oggi,noi}, we conclude that the interference effect hinders a determination of the hierarchy at the Daya Bay II reactor experiment locations proposed in Refs.~\cite{caojunseminario,yifangseminario,weihai}.

On the other hand, these locations benefit from the flux of an additional reactor complex which may be approved and eventually built at HuiDong.  As we will see below, the additional flux would suffice to make these the ideal sites (along with a site near DongKeng located between the TaiShan and YangJiang reactor complexes) except for the fact that we know of no mountain at a medium baseline which is equidistant from Daya Bay and HuiDong (TaiShan and YangJiang in the case of the DongKeng site).  The maximum elevation at an equidistant site is only about 200-300 meters in both cases, meaning that backgrounds caused by atmospheric muons may significantly reduce the detector efficiency.  Both are about 3 km from 500 meter mountains, but we will see that this 3 km baseline difference leads to another interference effect which makes experiments under these mountains essentially insensitive to the hierarchy, unless the direction in which a neutrino traveled can be sufficiently well established to determine from which reactor it originated with sufficiently high probability,  for example if at least in some energy range one can improve upon the technique used by the CHOOZ collaboration in Ref.~\cite{choozdirezione}.

We will therefore find that the best locations are perpendicular to the Daya Bay and Ling Ao reactors, where this interference effect is absent.  As there will be backgrounds from other reactors, for example, the TaiShan and YangJiang reactors which will begin to come into service next year, a short baseline is essential to optimize the signal to noise ratio.  The more reactors that will be built, the shorter the optimal baseline will be.  We have simulated various combinations of reactors and in general we find that the best location is the 1000 meter mountain BaiYunZhang which is 44.5 km from Daya Bay.  However a number of 700 meter mountains are located within 10 kilometers of BaiYunZhang.  These mountains offer somewhat larger baselines, which may be optimal if many of the planned reactors are not approved.

We begin in Sec.~\ref{montecarlo} with a description of our Monte Carlo and the way that we convert the observed neutrino spectrum into a determination of the neutrino mass hierarchy.  We define a number of quantities which reflect features of the Fourier transform of the spectrum, generalizing those of Ref.~\cite{caojun}, and then the hierarchy is determined by a linear combination of these quantities with weights that are optimized by a neural network.  It was noted in Ref.~\cite{oggi} that the features of a Fourier transform are very sensitive to the reactor spectrum and to $\mn32$.  It was shown in Ref.~\cite{noispurioso} that this dependence yields no information about the hierarchy but can be eliminated using a weighted Fourier transform.  We let the first layer of the neural network optimize these weights.  In Sec.~\ref{optsez} we systematically study reactors at different baselines from single detectors, finding the optimal baseline to be in the 48-52 km range in this case.  We also study the interference effect at various angles.  Finally in Sec.~\ref{rissez} we describe the detector and reactor locations considered and present the results of our simulations.  We find that even reactors 200 km away from the detector can appreciably impede the determination of the hierarchy so that, as a high signal to background ratio becomes essential, they reduce the optimal baseline from that found in Sec.~\ref{optsez}.  For simplicity and an easy comparison with earlier results, we do not use the neural network in the analysis in Sec.~\ref{rissez} but instead consider the analysis method of Ref.~\cite{caojun}, although an application of our neural network to BaiYunZhang is described in Subsec.~\ref{baiyunsez}.

After this paper was written we received the preprint~\cite{petcov2012} which presents a $\chi^2$ analysis of the spectrum of a single reactor at a baseline of 60 km.  That study differs from ours in that it does not include statistical errors, thus as may be expected it found that about an order of magnitude less flux is required for a determination of the hierarchy with comparable confidence.  Our results for single reactor experiments are, on the other hand, broadly compatible with the Fourier transform based analysis of Ref.~\cite{caojunseminario} and with the $\chi^2$ analysis of Ref.~\cite{oggi}.

\section{The Monte Carlo} \label{montecarlo}

\subsection{The Parameters}

This paper builds upon the simulations of medium baseline neutrino oscillation in Refs.~\cite{caojun2,caojunseminario}.  For ease of comparing with the results of~\cite{caojunseminario}, we have adopted the same values for the neutrino mass matrix parameters
\beq
\m21=7.59\times 10^{-5}\rm{\ eV}^2\hsp
\mn32=2.43\times 10^{-3}\rm{\ eV}^2\hsp
\spp2212=0.861.
\eeq
In particular we have used Daya Bay's March result~\cite{dayabay}
\beq
\spp2213=0.092
\eeq
and not their updated result~\cite{neut2012}.  Similarly we have assumed an absolute detector energy resolution
\beq
\sigma_E=.03\sqrt{E_ {pr}\rm{(MeV)}}
\eeq
where the prompt energy $E_{pr}$ is related to the neutrino energy $E$ and the positron energy $E_e$ by
\beq
E_{pr}=E_e+m_e=E-m_n+m_p+m_e\sim E-0.8\ \mathrm{MeV}.
\eeq

\subsection{The Simulation}

The overall reactor flux normalization is fixed to be that of Ref.~\cite{caojunseminario}.  It is asserted that the Daya Bay plus Ling Ao I and II reactor complexes, which together have 17.4 GW of thermal capacity, lead to the observation of 25,000 neutrinos at a 20 kton detector with a baseline of 58 km.  To be precise this does not fix the reactor flux normalization, but rather the overall normalization of the product of the reactor flux, the antineutrino cross section of the target, the detector efficiency and the effect of neutrino oscillations.  This condition, together with the tree level inverse $\beta$ decay cross section in Ref.~\cite{sezionedurto}
\beq
\sigma(E)=0.0952\times 10^{-42}\mathrm{cm}^2(E_e\sqrt{E_e^2-m_e^2}/\mathrm{MeV}^2)
\eeq
and the oscillation probability $P_{ee}$
\bea
P_{ee}&=&\sp413+\cp412\cp413+\sp412\cp413+\frac{1}{2}(P_{12}+P_{13}+P_ {23})\nonumber\\
P_{12}&=&\spp2212\cp413\cos\left(\frac{\m21L}{2E}\right)\hsp
P_{13}=\cp212\spp2213\cos\left(\frac{\mn31L}{2E}\right)\nonumber\\
P_{23}&=&\sp212\spp2213\cos\left(\frac{\mn32L}{2E}\right)\nonumber
\eea
are then used to normalize the effective reactor flux $\Phi(E)$ per time per GW of thermal capacity.  This reactor flux is effective in the sense that it is already multiplied by the efficiency of the detector.

With this optimistic normalization in hand, the average number density of antineutrinos at energy $E$ from a given reactor which would be observed during 3 years at a 20 kton detector at a baseline $L$ is then
\beq
N_{\rm{th}}(E,L)=\frac{\Phi(E)\sigma(E)P_{ee}(L/E)}{4\pi L^2}.
\eeq
In each 60 kton year experiment we simulate
\beq
N_{\rm{th}}(L)=\int dE N_{\rm{th}}(E,L)
\eeq
neutrinos from each reactor.  Thus our simulations differ from a real experiment in that the total neutrino flux normalization is fixed and is known precisely.  We will recreate 120  year experiments by simply summing the neutrinos from pairs of 60 kton year experiments.  To minimize the relative statistical errors between experiments with different exposure times, 120 kton year experiments will be created from 60 kton experiments reported on the same tables.


As can be seen in Table~\ref{disttab}, unlike Refs.~\cite{caojun2,caojunseminario} our total neutrino flux in each experiment depends on $L$.  In fact it increases faster than $1/L^2$ as $L$ is decreased from 58 km because there is less loss due to 1-2 neutrino oscillations.  This means that at short baselines, such as 40-50 km, our simulations allow a precise determination of the high energy peaks, which are hardly depleted by 1-2 oscillation.  The energies $2\pi/\Delta M^2_{\rm{eff}}$  of these peaks determine the effective mass splitting~\cite{parke2005,noi}
\beq
\Delta M^2_{\rm{eff}}=\cp212\mn31+\sp212\mn32. \label{meff}
\eeq
As a result we will see that our simulations favor shorter baselines than those preferred in Refs~\cite{caojun2,caojunseminario}.  We will also see that shorter baselines are preferred as they reduce the fractional backgrounds from distant reactors, backgrounds not included in previous studies.

Once we have fixed the numbers and energies of the neutrinos arriving at our detector, the finite resolution $\sigma_E$ of the detector is applied.   With infinite luminosity the observed neutrino spectrum $P_{obs}(E)$ would be the theoretical spectrum convoluted with a Gaussian of width $\sigma_E$.  To implement this effect at a finite luminosity we shift the energy of a neutrino by a random variable  $\delta$ with Gaussian probability density $\rm{exp}(-\delta^2/2\sigma_E^2)/\sqrt{2\pi\sigma_E}$.

Thus our simulation correctly accounts for statistical errors in the determination of the energy due to the finite number of photoelectrons detected.  However it does not take into account a systematic nonlinear error in the determination of energy.  While linear errors are harmless to the Fourier analysis that will be performed later, they simply shift the points, even a small nonlinear uncertainty can alter~\cite{petcov2010} or destroy~\cite{oggi} the effectiveness of such an analysis.  In practice the energy can be calibrated, at least at some energies and at some points inside of the detector, by inserting radioactive sources.  A peak by peak analysis of the spectrum can be performed just using peaks at these energies~\cite{noi} but this requires more flux than a Fourier analysis.

\subsection{Improvements over Previous Simulations}

Although the simulation does not account for the unknown nonlinear response of the detector, it does include an interference effect~\cite{noi} which has so far not been simulated in the literature.  The reactor complex at Daya Bay is separated by more than a kilometer from the reactor complexes Ling Ao I and Ling Ao II.  While this set of reactors is in general an ideal choice for medium baseline experiments, given the high flux and the existence of 8 short baseline detectors, in general the 1-3 oscillations from Ling Ao and Daya Bay will be out of phase at low energies.  The measurement of 1-3 oscillations at these low energies are essential to break a degeneracy which prohibits the determination of the neutrino mass hierarchy~\cite{parke2007,oggi,noi}.  Thus such interference poses a severe problem for the determination of the hierarchy.  However, since the Daya Bay, Ling Ao I and Ling Ao II complexes, like RENO, lie along a line, it is possible to evade this interference effect if the detector is roughly orthogonal to the line.  The price of this choice is that one cannot benefit from the flux at the proposed HuiDong reactor complex, which would no longer be equidistant from the detector.

Our simulation also includes the flux from many distant reactors, of which TaiShan and YangJiang will begin commercial operations next year.  We will see that these backgrounds have a nonnegligible effect on the probability of determination of the neutrino mass hierarchy and as a result they decrease the optimal baseline.   They provide more background but also potentially more usable information for the determination of $\theta_{12}$.

\section{Data Analysis}

\subsection{Fourier Analysis and Reactor Flux Models} \label{flusssez}

In this note we will not consider a peak energy analysis~\cite{noi} of the data, although this requires an understanding of the nonlinear response of the detector only at the locations of any two peaks, where it may be calibrated using radioactive decays.  Instead we will consider a Fourier analysis~\cite{hawaii}, which has the advantage that it combines neutrinos from multiple peaks and so requires less flux, despite the fact that it requires an unprecedented understanding of the nonlinear response of the detector~\cite{oggi} over a wide range of energies. Following Refs.~\cite{caojun,caojun2} we will decompose this transform into a real and a complex part
\bea
F^i_c(k)&=&\sum_j w^i(E_j) N(E_j)  \cos\left(\frac{kL}{E_j}\right)\nonumber\\
F^i_s(k)&=&\sum_j w^i(E_j) N(E_j)  \sin\left(\frac{kL}{E_j}\right) \label{fcos}
\eea
where we have summed over energy bins $j$, each centered at energy $E_j$ with $N(E_j)$ neutrinos observed.  We use 1 keV bins, but we have found that for such small bins our results are independent of the bin size.  The index $i$ labels different Fourier transforms obtained using different weights $w^i(E)$.   We will now explain the energy-dependent weight $w^i(E)$, which was introduced in Ref.~\cite{noispurioso}.

We will be interested in the form of the Fourier transforms near the 1-3 oscillation peak $k=\mn31/2$.  At these wavenumbers it is commonly believed that gross features of the spectrum are independent of the reactor flux and $P_{12}$, but in the case of a trivial weighting $w(E)=1$ this is not quite true.  Recently Ref.~\cite{oggi} found that the quantities (\ref{rlpv}) used in the analysis of Refs.~\cite{caojun,caojun2} depend strongly upon the reactor flux model used and for some flux models even upon $\mn32$, with variations as large as a factor of 5.  In Ref.~\cite{noispurioso} it was shown that the dependence upon the reactor flux model and $\mn32$ results from a dependence of the unweighted Fourier transform upon the high energy tail ($E>8$\ MeV) of the neutrino spectrum.  As the high energy spectrum is only sensitive to $\meff$ and therefore not to the hierarchy~\cite{parke2007,noi}, this dependence introduces a kind of nuisance parameter, impeding the discovery of the hierarchy using the unweighted transform.  In Ref.~\cite{noispurioso} it was shown that there exist weight functions $w(E)$ which gradually cut off the sensitivity to the high energy spectrum and so lead to a weighted Fourier transform without this spurious dependence.  Furthermore it was shown that the weight functions actually {\it{improve}} the sensitivity of the Fourier transform to the hierarchy.

Once the high energy dependence is eliminated, the sensitivity of the weighted Fourier transform to the reactor flux model becomes negligible.  Therefore, to maximize the compatibility of our analysis with that of Refs.~\cite{caojun,caojun2}, we will use the Gaussian fit to the old reactor flux model of Ref.~\cite{quadflusso}.  We will consider a general weight function
\beq
w^i(E)=a_1^{\ i} e^{-0.08\left(\frac{E}{\rm{MeV}}\right)^2}+a_2^{\ i}e^{-0.08\left(\frac{E}{\rm{MeV}}-3.6\right)^2}+a_3^{\ i}e^{-\frac{E}{3\ \rm{MeV}}}+a_4^{\ i}e^{-\frac{E}{6\ \rm{MeV}}}+a_5^{\ i}e^{-\left(\frac{E}{\rm{MeV}}-5.25\right)^2/4}. \label{peso}
\eeq
For the $i$th observable, the coefficients $a_j^{\ i}$ will be optimized by a neural network as described in Subsec.~\ref{neursez}.  The weight functions have been chosen so as to fall off fast enough at high energy that the spurious effect of high energy neutrinos on the determination of the hierarchy will be negligible.  They have also been chosen so as to allow the neural network to dynamically determine which part of the spectrum is the most important for a given observable.  However the basis of weights itself has not been optimized.  Once the basis has been optimized for a given experimental configuration, the chance of successfully determining the hierarchy will improve.

\subsection{Observables} \label{obssez}

Now that we have inserted a scale factor into the Fourier transform, the qualitative features of the transformed spectrum depend primarily upon $P_{13}$, which gives a symmetric $F_c(k)$ with a central peak and damped oscillations with a wavenumber of order $L\langle 1/E\rangle$ where $\langle 1/E\rangle$ is the average value of $1/E$.  At medium baselines, of order 40-80 km, this wavenumber is of order $\m21$.  Similarly the $P_{13}$ contribution to the sine transform $F_c(k)$ is antisymmetric about $k=\mn31/2$, with a maximum at higher $k$ and a minimum at lower $k$.  Again, as one varies $k$ from $k=\mn31/2$ one finds a series of ever shrinking oscillations separated by a characteristic distance of order $\m21$.  At $L=58$\ km analytic approximations to these features are derived in Ref.~\cite{noi}.

As the Fourier transforms are linear, they add the transform of the $P_{23}$ oscillations to that of the larger $P_{13}$ oscillations.  The hierarchy can be determined from the way in which the resulting asymmetric perturbation breaks the (anti)symmetry of the (sine) cosine transform of $P_{13}$.  It was observed in Ref.~\cite{caojun} and derived in Ref.~\cite{noi} that in the case of the normal (inverted) hierarchy the $P_{23}$ oscillations render the first minimum $R$ to the right of the global maximum of $F_c$ larger (smaller) than its mirror image $L$ and similarly render the maximum $P$ of $F_{s}$ just to the right of $k=\mn31/2$ larger (smaller) than the minimum just to its left $V$.    Ref~\cite{caojun} introduced two parameters which characterize these effects
\beq
p_1=RL=\frac{R-L}{R+L}\hsp
p_2=PV=\frac{P-V}{P+V} \label{rlpv}
\eeq
finding that positive (negative) values of these two parameters tend to indicate the normal (inverted) hierarchy.  We introduce the additional notation $p_1$ and $p_2$ to simplify the equations that follow.

In Ref.~\cite{noi} two more parameters were suggested.  First, the value $\phi$ of $F_s$ at the maximum of $F_c$, which is positive (negative) for the normal (inverted) hierarchy.  Second, the difference in values $m_\pm$ of the maxima of hierarchy-dependent nonlinear Fourier transforms
\beq
F_n^\pm(k)=\sum_j w^i(E_j) N(E_j)  \cos\left(k\frac{L}{E_j}\pm 2\pi\alpha\left(\frac{k}{2\pi}\frac{L}{E_j}\right)\right).
\eeq
These can be encoded into the normalized observables
\beq
p_3=\frac{\phi}{P+V}\hsp
p_4=\frac{m_+-m_-}{m_-+m_+}.
\eeq
We have repeated our analysis using a new parameter $p_5$ which is defined identically to $p_3$ except that it is evaluated at the maximum of the norm ($F_c^2+F_s^2$) of the Fourier transform and not at the maximum of $F_c$ alone.  The resulting improvement in our determination of the hierarchy is within our statistical errors and so we have not included it in our results below.

\subsection{Neural Network} \label{neursez}

In Subsec.~\ref{obssez} we identified four quantities $p_i$ which may be used to determine the hierarchy.  The sign of any one  is already a good indicator of the hierarchy, in fact they are reasonably degenerate.

Our goal is to determine the best indicator, that which, given a dataset, has the highest probability of correctly determining the mass hierarchy.  As these indicators are not precisely degenerate, the best indicator will not be a single $p_i$ but rather some combination of all 4.  For simplicity we will consider linear combinations
\beq
I=b_0+\sum_{i=1}^{4}b_i p_i \label{secondo}
\eeq
where $b_i$ are 5 constants.   To each of these observables $p_i$ except for the constant $i=0$ we associate a weight function $w^i(E)$ which is determined by the 5 constants $a_j^{\ i}$ defined in Eq.~(\ref{peso}).   A different weight may be optimal for each observable $p_i$, thus in all we need to optimize 20 constants $a_j^{\ i}$ and 5 constants $b_i$.  The optimal values are those such that, given a set of experiments in which the neutrino mass hierarchy is normal in as many as it is inverted\footnote{This corresponds to a Bayesian prior which assigns a 50\% chance to each hierarchy.  To use the hierarchy confidence that may be obtained at NO$\nu$A or potentially T2K one need only match the ratio of hierarchies in the simulated experiments.}, the chance of success is the highest.  The chance of success is defined to be the average of the percentage of normal hierarchy experiments for which $I$ is positive with the number of inverted hierarchy experiments for which $I$ is negative.  Clearly the overall scale of $b_i$ is irrelevant, but the optimal choice of $b_i$ in general depends on the baseline, the reactors that are operational, the geometry of the reactors with respect to the detector and even the time for which the experiment has run.

We will optimize these coefficients using a 2 layer neural network.  The first layer, for each $p_i$, will optimize the weight coefficients $a_j^{\ i}$ so as to maximize the number of experiments in which $p_i$ alone determines the hierarchy.    The second layer, as described above, then optimizes the weights $b_i$ of the weighted $p_i$'s.  As we will have fit 44 coefficients one may worry about overfitting if the number of experiments is too small.  Therefore we have only applied the full 2 layer neural network to BaiYunZhang, as that is the only site for which we have simulated 4000 experiments.  For this site overfitting effects appear to affect the probability of success by less than 1\%.   To be sure that our numbers are not inflated by overfitting, in every case in which we quote the result of a neural network the data used to test the chance of success has no overlap with the data used to train the neural network.


Summarizing, for each value of $i$ from 1 to 4 the first layer of the neural network optimizes $a_j^{\ i}$ of Eq. (\ref{peso}) so as to maximize the probability of success of $p_i$ alone.  Then the second layer optimizes the $b_i$ in Eq. (\ref{secondo}) so as to maximize the probability of success of the indicator $I$.

\subsection{The Neural Network Applied to BaiYunZhang} \label{baiyunsez}

\begin{table}[position specifier]
\centering
\begin{tabular}{c|l|l|l|l|l|l}
Weight&$p_1$&$p_2$&$p_3$&$p_4$&$p_1+p_2$&NN Level 2\\
\hline\hline
1&74.6\%&76.9\%&74.9\%&73.0\%&77.5\%&\\
\hline
$e^{-0.08\left(\frac{E}{\rm{MeV}}\right)^2}$&75.7\%&77.0\%&74.5\%&72.6\%&77.6\%&\\
\hline
$e^{-0.08\left(\frac{E}{\rm{MeV}}-3.6\right)^2}$&74.9\%&77.0\%&74.8\%&73.0\%&77.3\%&\\
\hline
$e^{-E/(3\ \rm{MeV})}$&75.5\%&74.1\%&71.6\%&69.4\%&75.8\%&\\
\hline
$e^{-E/(6\ \rm{MeV})}$&76.2\%&76.1\%&73.4\%&71.4\%&71.6\%&\\
\hline
$e^{-\left(\frac{E}{\rm{MeV}}-5.25\right)^2/4}$&70.8\%&74.2\%&72.4\%&77.4\%&72.3\%&\\
\hline
NN level 1&75.7\%&77.6\%&76.3\%&77.6\%&&78.0\%\\
\hline
\end{tabular}
\caption{The probability of successfully determining the hierarchy at a 20 kton detector under BaiYunZhang after 3 years of running using various indicators and layers of the neural network independently.  It is assumed that all planned reactors have been built.  The bottom right percentage is our final answer, it does not suffer from overfitting as the neural network is not trained on the same data to which it is applied.  The other entries are the average chances of success of 2000 experiments with each hierarchy.}
\label{neurtab}
\end{table}

We will now illustrate the functioning of the neural network on 60 kton years of neutrino detection underneath the mountain BaiYunZhang, described in Table~\ref{rivelatoretab}.  We will assume that all of the reactors described in Table~\ref{reactortab} are operational.   This is an overly pessimistic  assumption, as Daya Bay II is scheduled for completion in 2020~\cite{yifangseminario} whereas the reactors at HuiDong and Lufeng are still awaiting final approval.  Once construction begins, an individual reactor generally takes about four and a half years to build in China, and generally at a complex of reactors, construction on one reactor begins each year.   Therefore, unless this scheduling changes, one may expect HuiDong and LuFeng to take 10 years from the beginning of construction to reach full capacity, and thus not to be at full capacity before data taking begins at Daya Bay II.

We have considered four sets of 1000 experiments, each of which contains 500 with each hierarchy.  The neural network is trained separately on each subset of three sets and then tested on the remaining set.  The final result for the probability of success, displayed in the bottom right corner of Table~\ref{neurtab} is the average of the four probabilities of success obtained from the testing of the neural network on each set.   Notice that while the nonlinear Fourier transform $p_4$ and the high energy weight $e^{-\left(\frac{E}{\rm{MeV}}-5.25\right)^2/4}$ in general perform poorly, the high energy nonlinear transform alone produces a $77.4\%$ chance of success, which is the highest of any individual observable with any individual weight.  Thus the nonlinear Fourier transform performs better at high energies, whereas the linear Fourier transform performs better at low energies.  As the oscillation probability depends on $L/E$, this will imply that the nonlinear transform is optimized at shorter baselines than the linear transform methods, which is why we obtain a shorter optimal baseline than was obtained in Refs.~\cite{caojun2,caojunseminario}.

While the neural network does outperform $p_1+p_2$ with any weight, this effect is quite small.  When we turn to simulations with a single baseline in Subsec.~\ref{distsez} we will see that at such short baselines even the second layer of the neural network alone significantly outperforms $p_1+p_2$.  However this improvement is reduced when multiple baselines are present.  Of course an optimization of the basis of weights and a longer running of the neural network, trained on more data, will enhance its performance.


\section{Optimizing the Baseline and Geometry} \label{optsez}

\subsection{The Optimal Baseline} \label{distsez}

\begin{table}[position specifier]
\centering
\begin{tabular}{c|l|l|l}
Baseline&$N_{\overline{\nu}_e}$ at 60 ky&Hierarchy at 60 ky&Hierarchy at 120 ky\\
\hline\hline
42 km&64,860&86.0\%\ (81.9\%)&94.1\%\ (90.3\%)\\
\hline
44 km&55,578&86.4\%\ (83.3\%)&95.1\%\ (90.3\%)\\
\hline
46 km&48,030&87.7\%\ (84.9\%)&95.5\%\ (93.1\%)\\
\hline
48 km&41,900&88.4\%\ (86.2\%)&95.5\%\ (93.2\%)\\
\hline
50 km&36,931&88.7\%\ (87.1\%)&95.4\%\ (94.2\%)\\
\hline
52 km&32,915&88.1\%\ (86.4\%)&95.0\%\ (92.9\%)\\
\hline
54 km&29,679&86.4\%\ (85.8\%)&94.8\%\ (93.9\%)\\
\hline
56 km&27,080&86.8\%\ (85.5\%)&93.7\%\ (93.2\%)\\
\hline
58 km&25,000&84.3\%\ (84.0\%)&94.1\%\ (93.4\%)\\
\hline
60 km&23,342&81.9\%\ (81.8\%)&93.2\%\ (92.4\%)\\
\hline
\end{tabular}
\caption{The observed $\overline{\nu}_e$ flux from a single 17.4 GW reactor and the probability of determining the hierarchy as a function of baseline and number of kton years determined using the second layer of the neural network.  The neural networks trained on the neighboring baselines, and the values of the coefficients $b_i$ used to determine the chance of success are the averages of the coefficients at the available neighboring baselines.  Probabilities in parenthesis are obtained using $p_1+p_2$ as in Ref.~\cite{caojun}.}
\label{disttab}
\end{table}

As a first application of our simulation we have attempted to determine the optimal baseline in an idealized situation in which all 17.4 GW of thermal capacity are located at a single reactor, at a fixed baseline.  This eliminates the interference effects of Ref.~\cite{noi} which will be investigated in Subsec.~\ref{intsez}.  The probability of success then depends only upon the baseline, the number of ktons of the detector multiplied by the number of years of observation and the method of data analysis.  We will consider an unweighted Fourier transform analysis with two indicators, $p_1+p_2$ which was introduced in Ref.~\cite{caojun} and also the second layer of our neural network.  The results, considering 2000 experiments of each hierarchy, are displayed in Table~\ref{disttab}.

\subsection{Interference Between Reactors in the Same Complex} \label{intsez}

No single reactor produces enough flux to determine the neutrino hierarchy in a reasonable amount of time.  Thus it is inevitable that a complex of reactors needs to be used, and often the distances between these reactors is considerable.  For example in Ref.~\cite{weihai} the proposed site for Daya Bay II is 3.5 km closer to the reactor complexes at Daya Bay and Ling Ao than to that planned at HuiDong.  These different baselines mean that for some energies the neutrinos from one reactor will arrive at the 1-3 oscillation maximum while neutrinos from the other will arrive at the 1-3 oscillation minimum, greatly reducing the amplitude of the 1-3 oscillations whose observation is necessary to determine the hierarchy~\cite{noi}.  Note that the neutrinos arriving from different reactors are not coherent, the interference effect results from the addition of probabilities and not wavefunctions.

Such an interference effect is present using the reactors at Daya Bay and Ling Ao alone, because Daya Bay and Ling Ao I are separated by 1.1 km and Ling Ao II is 500 meters further.   Fortunately these reactors all lie more or less along a line, so a medium baseline detector perpendicular to this line will be the same distance from each reactor, eliminating the interference effect.  In Table~\ref{inttab} we have determined the effect of this interference on the probability of success for a detector as a function of its distance from the center of mass of Daya Bay and Ling Ao, its angle with respect to this line and the number kton years of observations.  To illustrate the effect of the angle, the interference from other, more distant, reactors is not considered.  These will be included in the full simulations reported in Sec.~\ref{rissez}.

\begin{table}[position specifier]
\centering
\begin{tabular}{c|l|l|l|l|l|l|l}
$L$&ky&$0^\circ$&$15^\circ$&$30^\circ$&$45^\circ$&$60^\circ$&$75^\circ$\\
\hline\hline
50 &60&66.3\ (66.4)&66.4\ (66.5)&70.6\ (70.7)&76.4\ (75.9)&82.0\ (81.5)&85.8\ (84.7)\\
\hline
50 &120&73.5\ (72.1)&72.1\ (71.8)&78.3\ (76.5)&85.1\ (84.0)&90.8\ (89.7)&93.9\ (91.7)\\
\hline
58 &60&63.6\ (64.2)&63.8\ (64.4)&67.9\ (67.6)&73.8\ (73.7)&79.8\ (78.9)&83.3\ (82.4)\\
\hline
58 &120&73.0\ (71.6)&74.2\ (72.4)&77.5\ (75.7)&84.9\ (83.5)&89.5\ (88.2)&93.0\ (91.5)\\
\hline
\end{tabular}
\caption{The probability (\%) of determining the hierarchy as a function of baseline in kilometers and number of kton years determined using unweighted Fourier transforms and only the second layer of the neural network.  The neural network is trained on the neighboring angles.  The value in parenthesis is the percentage chance of success with only $p_1+p_2$.  The baseline is the distance from the center of mass of the Daya Bay, Ling Ao I and II reactor complexes considering the distances between these complexes.  The angles are measured with respect to the line that nearly passes through all three complexes.}
\label{inttab}
\end{table}

\section{Comparing Detector Locations Near Daya Bay} \label{rissez}

\subsection{Locations of Reactors and Detectors}

China's Guangdong province is among the best locations for a medium baseline reactor experiment because it contains a powerful reactor complex consisting of 2 reactors at Daya Bay and 4 at Ling Ao and yet it is free from the large reactor neutrino backgrounds caused by many smaller complexes in France and, modulo tsunami induced shutdowns, in Japan.  In the next few years a number of new reactor complexes will be completed in Guangdong.  Two distant complexes, TaiShan and YangJiang, will see their first reactors generate power already next year.  The TaiShan reactors will be the world's first completed EPR reactors, which is an advantage as it will mean more than 50\% more neutrino flux per reactor than the other reactors in Guangdong.  On the other hand, it means that there may be some deviation between its spectrum and that measured at Daya Bay. Two closer complexes, HuiDong and LuFeng, have already passed several critical steps in the approval process and may be built.  The relevant reactors are listed in Table~\ref{reactortab}.

\begin{table}[position specifier]
\centering
\begin{tabular}{l|l|l|l|l}
Name&Status&Latitude&Longitude&Thermal Power\\
\hline\hline
Daya Bay&Operational&$22^\circ$ 35'  53" N&$114^\circ$ 32' 35" E& 5.8 GW\\
\hline
Ling Ao I&Operational&$22^\circ$ 36' 19" N& $114^\circ$  33' 4" E& 5.8 GW\\
\hline
Ling Ao II&Operational&$22^\circ$ 36' 31" N& $114^\circ$  33' 14" E& 5.8 GW\\
\hline
TaiShan I&Under Construction&$21^\circ$ 55' 9" N& $112^\circ$  58' 57" E& 9.2 GW\\
\hline
TaiShan II&Planned&$21^\circ$ 55'  N& $112^\circ$  59' E& 9.2 GW\\
\hline
YangJiang I&Under Construction&$21^\circ$ 42' 29" N& $112^\circ$  15' 32" E& 5.8 GW\\
\hline
YangJiang II&Under Construction&$21^\circ$ 42' 36" N& $112^\circ$  15' 41" E& 5.8 GW\\
\hline
YangJiang III&Planned&$21^\circ$ 43'  N& $112^\circ$  16'  E& 5.8 GW\\
\hline
HuiDong&Planned&$22^\circ$ 42'  N& $115^\circ$  0'  E& 17.4 GW\\
\hline
LuFeng&Planned&$22^\circ$ 45'  N& $115^\circ$  49'  E& 17.4 GW\\
\hline
\end{tabular}
\caption{Reactors in Guangdong}
\label{reactortab}
\end{table}

Depending on the detector location the new reactors may help or hinder the determination of the neutrino hierarchy.  They will help at each detector which is equidistant from two reactors,  as the fluxes add without interference.  This is the case for the proposed location of Ref.~\cite{yifangseminario} and also at our proposed location DongKeng.  If the two reactors are nearly at the same distance, as in the proposal of Ref.~\cite{weihai} and our proposed location HuangDeDing, the addition of the second reactor will increase the flux but interference between the neutrinos from different reactor complexes will decrease the amplitudes of the 1-3 oscillations.

However in general the additional reactors will be so distant that the 1-3 oscillation peaks will be too close together to be distinguishable by a detector with resolution $\sigma_E$.  As a result, the additional reactors will simply supply a background which impedes the measurement of the hierarchy.

A medium baseline detector experiment may also be used to measure $\theta_{12}$.  This can be done by comparing the flux at the $1-2$ oscillation minima and maxima.  If the new reactor is twice as far away as the desired reactor, as is the case for most of the positions that we will consider, then the distant reactor will be at its 1-2 minimum at the same energy range in which neutrinos from the near reactor arrive at their 1-2 maximum.  This means that at the 1-2 maximum, where sensitivity to the hierarchy and to $\theta_{12}$ is maximized, the neutrino flux will be dominated by noise from the distant reactor~\cite{noi}.  In general this makes the determination of both the hierarchy and $\theta_{12}$ more difficult.  However the 1-2 oscillations of distant reactors do lie within the resolution of the detector and so these can be used to gain more information about $\theta_{12}$.  Furthermore, if multiple detectors are considered, then they can break the degeneracy between $\theta_{12}$ and flux from distant reactors, allowing a more precise determination of both.

\begin{table}[position specifier]
\centering
\begin{tabular}{l|l|l|l|l|l|l|l|l}
Name&Altitude&Latitude (N) &Longitude (E) &DB&TS&YJ&HD&LF\\
\hline\hline
BaiYunZhang&1000 m&$22^\circ$ 53'  52"&$114^\circ$ 15' 14"& 44.5&170.5&244.1&79.5&161.6\\
\hline
ShiYaTou&500 m&$22^\circ$ 52'  14"&$114^\circ$ 17' 28"& 39.7&171.6&245.7&75.5&157.6\\
\hline
ShuangFeiJi&700 m&$22^\circ$ 54'  19"&$114^\circ$ 10' 0"& 51.6&164.2&237.0&88.3&170.6\\
\hline
SanJiaBi&600 m&$22^\circ$ 54'  8"&$114^\circ$ 10' 41"& 50.6&164.9&237.8&87.3&171.6\\
\hline
XiangTouShan&800 m&$23^\circ$ 15'  24"&$114^\circ$ 21' 0"& 75.4&205.0&275.3&90.4&160.9\\
\hline
BaiMianShi&400 m&$23^\circ$ 6'  27"&$114^\circ$ 37' 2"&  56.8&214.1&288.0&60.3&129.7\\
\hline
DongKeng&200 m&$22^\circ$ 6'  4"&$112^\circ$ 31' 9"& 216.5&51.8&51.0&264.1&347.3 \\
\hline
HuangDeDing&500 m&$22^\circ$ 5'  23"&$112^\circ$ 29' 55"& 218.9 &53.3&48.8&266.5&349.7\\
\hline
\end{tabular}
\caption{Potential locations for Daya Bay II detectors and distances to reactors in km. DB~indicates the distance to the weighted center of Daya Bay and Ling Ao I and II.}
\label{rivelatoretab}
\end{table}

In Table~\ref{rivelatoretab} we list the detector positions that we propose together with BaiMianShi, which was proposed in Ref.~\cite{weihai}.  These points are chosen to be at a medium baseline, to minimize interference effects, to be underneath mountains to minimize backgrounds and when possible to be a similar distance from multiple reactors to enhance the useful neutrino flux.  No site that we have found simultaneously achieves all of these goals.

\subsection{Monte Carlo Results}

The simulation of planned reactors is impeded by the fact that we do not know the configurations of the reactors.  These configurations are important when the reactors are used as sources of flux because it determines the reduction in the resulting 1-3 peaks due to interference.  It is not important when the reactors provide a background source.  While in general these configurations cannot be predicted, one natural guess is that the reactors will follow the coast, which suggests for example that interference will not pose a problem for neutrinos generated at TaiShan.  However we have made no such assumptions in our analysis.  In Tables~\ref{risultatitab} and \ref{risultati2tab} we present our 60 kton year and 120 kton year results respectively.  We use a Roman numeral I to signify assumed interference at unbuilt reactors equal to that of Daya Bay and Ling Ao as observed at BaiMianShi and a Roman number II to signify no interference at unbuilt reactors.

\begin{table}[position specifier]
\centering
\begin{tabular}{l|l|l|l|l|l|l}
Name&Simulations&All&No LF&No HD&No HD/LF&DB and LA\\
\hline\hline
BaiYunZhang&4000&77.6\%&77.9\%&81.0\%&82.4\%&83.6\%\\
\hline
SYT+XTS&2000&71.5\%&73.0\%&75.8\%&76.4\%&78.3\%\\
\hline
SanJiaBi&2000&77.5\%&77.9\%&81.1\%&82.6\%&85.8\%\\
\hline
ShuangFeiJi&2000&74.6\%&76.7\%&80.3\%&81.7\%&85.2\%\\
\hline
BaiMianShi I&2000&53.0\%&51.7\%&65.3\%&68.1\%&72.2\%\\
\hline
BaiMianShi II&2000&49.3\%&48.9\%&64.6\%&66.7\%&69.8\%\\
\hline
BMS Ideal I&2000&74.4\%&78.3\%&62.9\%&66.5\%&68.9\%\\
\hline
BMS Ideal II&2000&86.4\%&88.2\%&62.9\%&67.2\%&70.3\%\\
\hline
DongKeng I&2000&74.9\%&74.5\%&74.6\%&75.2\%&49.3\% \\
\hline
DongKeng II&2000&87.8\%&87.9\%&88.2\%&88.4\%&50.8\% \\
\hline
HuangDeDing I&2000&58.5\%&58.7\%&58.8\%&59.2\%&50.1\%\\
\hline
HuangDeDing II&2000&63.7\%&63.6\%&65.0\%&64.4\%&50.8\%\\
\hline
\end{tabular}
\caption{Potential locations for Daya Bay II detectors and chances of success using $p_1+p_2$ for 60 kton years of flux and choices of active reactors.  The first column considers all planned reactors.  In the second LuFeng is excluded, in the third HuiDong is instead excluded, in the fourth both are excluded and in the last only Daya Bay and Ling Ao are considered.  SYT+XTS refers to two detectors, one under ShiYiTou and one under XiangTouShan, each of which is half as large as indicated in the corresponding column, simply summing the $L/E$ spectra.  BMS ideal is a 200 meter high point 2 km from BaiMianShi which is equidistant from the center of mass of the Daya Bay/Ling Ao complex and HuiDong. Both BaiMianShi I and BMS Ideal I assume that the baselines to the HuiDong reactors will differ by as much as those to Daya Bay and Ling Ao reactors, whereas II assume that the HuiDong reactors are coincident.  Similarly DongKeng and XiKengDing I (II) assume that there is (not) interference at TaiShan and YangJiang.}
\label{risultatitab}
\end{table}

Note that the sites with the highest percentage of success, BMS Ideal and DongKeng, have the smallest mountains.  This is because both are equidistant from two reactor complexes so as to benefit from the increased flux, but there is no peak at that location.  Instead they are under the highest equidistant point, which is between 200 and 300 meters in both cases.  The peaks of the corresponding mountains, BaiMianShi and XiKengDing, are each over 500 meters high but as can be seen from the tables, the interference effects are so great that these do not allow accurate determinations of the hierarchy.

Instead the best locations for a neutrino detector lie under the mountains orthogonal to the line between Daya Bay and Ling Ao.  A number of mountains between 44 and 51 km offer similar chances of determining the hierarchy.  In general the closest and highest, BaiYunZhang, is the best.  However if the planned reactors at HuiDong and LuFeng are not built, then the more distant mountains such as SanJiaBi and ShuangFeiJi gain a clear advantage.  Use of the full, 2-layer, neural network leads to a modest improvement in cases with large backgrounds with many reactors.  For example, if all proposed reactors are built, then with 60 kton years the full neural network yields a 78.0\% chance of success at BaiYunZhang and SanJiaBi and 76.5\% at ShuangFeiJi, representing on average a 1\% improvement over an analysis based on $p_1+p_2$ alone.  The statistical errors in these chances of success are slightly less than 1\%.  Indeed, one expects that with more data the chances of success at SanJiaBi and ShuangFeiJi will converge.

However even a 2$\sigma$ determination in Guangdong appears to require more than 120 kton years\footnote{The situation near the Yeonggwang reactor complex in South Korea is similar.  To avoid interference a detector must be to the south southeast.  But then one must choose between a 400 meter hill at a baseline of 47.4 km and Mudeungsan whose 950 meter peak is at 61.2 km.}.  As a result, either multiple detectors or a detector larger than 20 ktons may be preferable.  Multiple detectors are preferable for breaking the degeneracy between $\theta_{12}$ and the unknown neutrino flux from distant reactors~\cite{noi}.

\begin{table}[position specifier]
\centering
\begin{tabular}{l|l|l|l|l|l|l}
Name&Simulations&All&No LF&No HD&No HD/LF&DB and LA\\
\hline\hline
BaiYunZhang&2000&87.1\%&87.4\%&88.8\%&90.7\%&91.8\%\\
\hline
SYT+XTS&1000&80.3\%&81.9\%&84.0\%&85.2\%&87.3\%\\
\hline
SanJiaBi&1000&85.8\%&86.5\%&89.5\%&91.8\%&93.4\%\\
\hline
ShuangFeiJi&1000&84.4\%&85.6\%&88.7\%&91.2\%&94.5\%\\
\hline
BaiMianShi I&1000&55.3\%&55.7\%&73.7\%&78.8\%&80.7\%\\
\hline
BaiMianShi II&1000&44.9\%&45.7\%&74.0\%&77.3\%&80.8\%\\
\hline
BMS Ideal I&1000&84.9\%&88.3\%&72.4\%&78.1\%&79.2\%\\
\hline
BMS Ideal II&1000&93.9\%&95.3\%&71.4\%&79.4\%&82.4\%\\
\hline
DongKeng I&1000&83.8\%&83.3\%&84.0\%&83.7\%&49.7\% \\
\hline
DongKeng II&1000&95.0\%&95.6\%&95.3\%&95.8\%&51.5\% \\
\hline
HuangDeDing I&1000&62.3\%&61.3\%&63.4\%&62.7\%&51.5\%\\
\hline
HuangDeDing II&1000&72.7\%&71.8\%&72.7\%&72.6\%&50.6\%\\
\hline
\end{tabular}
\caption{Potential locations for Daya Bay II detectors and chances of success using $p_1+p_2$ for 120 kton years of flux and choices of active reactors. Labels as in Table~\ref{risultatitab}.}
\label{risultati2tab}
\end{table}

\section{Conclusions}

We have simulated medium baseline reactor neutrino experiments with a single baseline and also at real positions in Guangdong province.  We found that in the presence of a single reactor the optimal baseline is about 48-52 km.  The presence of multiple reactors and so multiple baselines has a number of implications.  First of all, if the difference between baselines is between 1 and 5 km then interference effects greatly reduce the chance of determining the hierarchy, although as the location BMS ideal shows, this loss can be more than offset by the presence of flux from an equidistant reactor complex.  Unfortunately in Guangdong it is difficult to find a mountain which is equidistant from a pair of reactor complexes at a sufficiently short baseline to take advantage of this.

We have also seen that the presence of multiple reactors, even at distances in excess of 200 km, significantly affects the chance of successfully determining the hierarchy at such an experiment.  In a few years a number of reactors about 200 km from both Daya Bay and RENO will become operational.  This loss can be somewhat compensated by reducing the baseline, thus adding flux from the desired reactors. As a result, the more reactors that will be built, the lower the optimal baseline.  This means that, in many cases the optimal location for a medium baseline reactor neutrino experiment is BaiYunZhang, only 44.5 km from Daya Bay.

The biggest challenge facing such an experiment is a nonlinear systematic error in the determination of the prompt energy ($E_e=E_\nu-0.8$\ MeV) from the number of photoelectrons observed in the photomultipliers.  A successful determination of the hierarchy using the Fourier transform requires the observed 1-3 peaks to be periodic in $L/E$, so that they add constructively in the Fourier transform.  Even a small systematic shift in the determination at the energy in some range of energies and at some locations in the detector can ruin this constructive interference, destroying the peak structure of the Fourier transform.  It was determined in Ref.~\cite{oggi} that to determine the hierarchy using Fourier transform methods one requires an understanding of the nonlinear response of the detector an order of magnitude better than that achieved at KamLAND.  In our study we have assumed that this nonlinear response is understood perfectly, and so our results yield upper bounds on the chance of successfully determining the hierarchy.

On the other hand the hierarchy can be determined by measuring the energies of just two peaks in the untransformed spectrum~\cite{noi}, one at high energy which determines $\meff$~\cite{parke2005,oggi,noi} and one at low energy which removes the remaining degeneracy in the mass differences.  The nonlinear response of the detector can be calibrated precisely at certain discrete energies using radioactive decays which produce neutrinos at those energies.  At such energies it is possible to understand the nonlinear response to a sufficient precision to determine the energy of the closest peaks, while at a sufficiently short baseline the peaks can simply be counted and identified to convert this energy into a combination of the mass differences as described in Ref.~\cite{noi}.  Therefore, with sufficient flux, an identification of individual peaks requires an understanding of the nonlinear detector response at only two energies, and not throughout the spectrum as is required by the Fourier transform method and in general the $\chi^2$ method.  A precise determination of the hierarchy will rely on a combination of these methods and so a determination of the optimal baseline and detector location needs to consider them all.

\section* {Acknowledgement}

\noindent
JE is supported by the Chinese Academy of Sciences
Fellowship for Young International Scientists grant number
2010Y2JA01. EC and XZ are supported in part by the NSF of
China.


\end {document}

10 years ago Petcov and Piai suggested that a 20-25 km baseline neutrino detector may be able to determine the neutrino mass hierarchy if the solar mass splitting $\m21$ is greater than about $10^{-4}\ \mathrm{eV}^2$~\cite{petcovidea}.  It turned out that $\m21$ is smaller than this threshold, meaning that the difference between the normal and inverted hierarchy cannot be seen at a baseline below about 40 km.  The low fluxes at these long baselines led various groups~\cite{hawaii,caojun} to conclude that the individual peaks in the neutrino spectrum would be difficult to resolve, and so the only hope for a hierarchy determination would be to sum them via a Fourier transform.  Even in this case the necessary detectors were extremely large and the experiments slow.

This all changed with the recent determination of $\theta_{13}$~\cite{daya,reno} confirming hints last year~\cite{doublechooz,globale} that $\spp2213$ is up to 10 times higher than had been considered by the authors of Refs.~\cite{hawaii,caojun}.  The larger mixing angle increases the sizes of the oscillations used to determine the hierarchy by an order of magnitude.  This means both that a reactor experiment to determine the hierarchy is now practical and also that the analysis of the optimal baseline and experimental configuration must now be redone.  Indeed, such a reanalysis is urgent as such an experiment will be built soon~\cite{caojunseminario,renonuturn,yifangseminario}.

We will perform this updated analysis in two papers, mirroring the structure of Ref.~\cite{caojun}.  While the second paper will include a description of an the results of our simulations, indicating for example the optimal baseline, in this first paper we will analyze a medium baseline reactor experiment analytically.  We will derive old observations relating observables of the Fourier transformed spectrum to the neutrino mass hierarchy and we will also provide new methods of determining various combinations of the neutrino masses and the hierarchy.  The optimal method will be a combination of the old and the new to be determined by simulations.  We will also discuss problems which have so far escaped attention in the literature.

In particular we will discuss the fact that, since reactors are typically separated by of order 1 km in a reactor array, the baselines of neutrinos from different reactors are different.  As a result the oscillations of low energy neutrinos, which perform half an oscillation traveling from one reactor to the next, will be invisible at the detector as the neutrinos from one reactor arrive at the oscillation minimum while the other is at its maximum.  We will show that, due to a new degeneracy between the hierarchy and an effective mass difference, a determination of the neutrino mass hierarchy is impossible in an experiment which cannot resolve these low energy peaks.  Thus the angle between the detector and the lines extending between reactors is an essential variable in the determination of the optimal detector location.  For example, for a linear array of reactors like RENO or Daya Bay plus Ling Ao, this effect can be eliminated if the detector is placed orthogonal to the array.

The greatest sensitivity to the hierarchy arises near the 1-2 oscillation minimum, at about 60 km.  However the flux from a distance reactor at the 1-2 oscillation maximum of 120 km will dominate over the flux of the near reactor in this region.  In fact, this is will be the case for a detector placed 60 km away from Daya Bay and Ling Ao in the orthogonal direction, as the proposed Haifeng reactor will be near the 1-2 maximum.  Also if a detector is placed equidistant from Daya Bay and Haifeng at the position suggested in Ref.~\cite{caojunseminario} then the 1-2 minimum neutrinos from Daya Bay and Haifeng will correspond to the 1-2 maximum for neutrinos from the proposed reactor at Lufeng~\cite{yifangseminario}.  This not only makes the determination of the hierarchy more difficult, but also is detrimental to the sensitivity to $\theta_{12}$.  The ideal solution to this problem is to use two detectors at different distances, say 40 and 70 km.  However a more economical solution is to keep the baselines short so that the flux from the desired reactor complex dominates over the fluxes from others, for example one can consider a baseline of 45-50 km.

We will begin in Sec.~\ref{teorsez} with a review of standard results on 3-flavor neutrino oscillations and the electron antineutrino survival probability.  We describe the interference between the 1-3 and 2-3 oscillations~\cite{petcovidea} which leads to beats and we numerically find the combined peaks.  Then in Sec.~\ref{masssez} we describe how the positions of various peaks can be used to obtain various combinations of the neutrino mass differences.  Each peak provides a different combination.  And we will see that the first 10 peaks do not allow a determination of the mass hierarchy, but the next 5 do, which is why short baselines are not sufficient for a determination of the hierarchy.  In Sec.~\ref{vincsez} we discuss the consequences of the finite energy resolution and the finite neutrino flux.  Both are provide obstacles to locating the $n>10$ peaks, and so for determining the hierarchy.  In Sec.~\ref{fouriersez} we discuss analyses of the Fourier transform of the neutrino spectrum.  These have the advantage that, when the nonlinearity of the detector response is well understood, they sum the peaks together, and so render the signal stronger. We rederive three old ways in which the hierarchy can be determined from this transformed spectrum and add two new methods to the list.

Finally in Sec.~\ref{intsez} we discuss the consequences of the fact that not all of the neutrinos detected traveled the same distance.  The distances may differ by of order a kilometer because the individual reactors in an array are not coincident, leading to an interference effect which greatly diminishes the amplitudes of the low energy, high $n$, peaks. We will see that this interference can be avoided if the detector is placed perpendicular to the array.   Also, while it has long been known~\cite{hawaii} that neutrinos from reactors at the 1-2 oscillation minimum baseline are the most useful for determining the hierarchy, we will see that this signal can be overwhelmed by neutrinos from distant reactors at the 1-2 maximum, and we see that this is indeed the case if a detector is placed at the 1-2 minimum orthogonal to the Daya Bay, Ling Ao reactor array.  This problem, as well as the related error in a determination of $\theta_{12}$, can be reduced by shortening the baseline or, if possible, adding another detector at a different baseline.

\section{The electron neutrino survival probability} \label{teorsez}

\subsection{Short and long oscillations}

The electron neutrino weak interaction eigenstate $|\nu_e\rangle$ is not an energy eigenstate $|k\rangle$, but it can be decomposed into a real sum of energy eigenstates
\beq
|\nu_e\rangle=\c12\c23|1\rangle+\s12\c13|2\rangle+\s13|3\rangle.
\eeq
In the relativistic limit, after traveling a distance $L$, the survival probability of a coherent electron (anti)neutrino wavepacket with energy $E$ can be expressed in terms of the mass matrix $\mathbf{M}$
\bea
P_{ee}&=&|\langle\nu_e|\mathrm{exp}\left(i\frac{\mathbf{M}^2L}{2E}\right)|\nu_e\rangle|^2\label{pee}\\
&=&\sp413+\cp412\cp413+\sp412\cp413+\frac{1}{2}(P_{12}+P_{13}+P_ {23})\nonumber\\
P_{12}&=&\spp2212\cp413\cos\left(\frac{\m21L}{2E}\right)\hsp
P_{13}=\cp212\spp2213\cos\left(\frac{\mn31L}{2E}\right)\nonumber\\
P_{23}&=&\sp212\spp2213\cos\left(\frac{\mn32L}{2E}\right)\nonumber
\eea
where $\m{i}{j}$ is the mass squared difference of mass eigenstates $i$ and $j$.  Notice that the survival probability is a sum of cosines and so its cosine Fourier transform with respect to the variable $L/E$ is just a sum of delta functions whereas its sine transform vanishes.

The three cosines in the survival probability (\ref{pee}) identify two characteristic frequencies of the $L/E$ oscillations.  The $P_{12}$ term oscillates at a low frequency
\beq
\frac{\m21}{2}\sim 3.8\times 10^{-5}\ \mathrm{eV}^2\sim 0.17\ \mathrm{MeV/km} .
\eeq
Therefore the maximum 1-2 oscillation occurs at
\beq
\frac{L}{E}=\frac{\pi}{\m21/2}\sim 18 \ \mathrm{km/MeV}. \label{unodue}
\eeq
A medium baseline reactor experiment, with a baseline of under 100 km,  can observe at most one or two such oscillations.  Instead such experiments will focus on shorter oscillations, characterized by the $P_{13}$ term, which have frequency
\beq
\frac{\mn31}{2}\sim 1.2\times 10^{-3}\ \mathrm{eV}^2\sim 5.5\ \mathrm{MeV/km}
\eeq
corresponding to a wavelength of
\beq
\Delta\left(\frac{L}{E}\right)=\frac{2\pi}{\mn31/2}\sim 1.1 \ \mathrm{km/MeV}. \label{corto}
\eeq
At a medium baseline reactor one may hope to see 5 to 15 such oscillations.

What about the $P_{23}$ term  in (\ref{pee})?  The frequency is $\mn32/2$, which is about 3\% more or less than $\mn31/2$ depending on the hierarchy.  However the amplitude is less than that of $P_{13}$ by a factor of
\beq
a=\tp212\sim 0.5.
\eeq
As the frequencies of the two short oscillations are similar but $P_{23}$ has a smaller amplitude, the total short distance oscillation $P_{13}+P_{23}$ is a deformation of $P_{13}$ alone.  However the 2-3 oscillations serve to slightly displace the 1-3 peaks and shift the amplitudes with a pattern which repeats at the beat frequency $\m12/2$.

More precisely, while the $n$th maximum of $P_{13}$ is at $L/E=4\pi n/\mn31$, the $n$th peak of $P_{13}+P_{23}$ is at
\beq
\frac{L}{E}=\frac{4\pi}{\mn31}(n+\an) \label{pichi}
\eeq
for a vector $\an$ which is determined entirely by the neutrino mass matrix.  As the derivatives of the neutrino spectrum and the 1-2 oscillations are small, the peaks of the total survival probability $P_{ee}$ and even its product with no-oscillation neutrino spectrum, which is the observed neutrino spectrum, are roughly located at the values given in Eq. (\ref{pichi}).

From Eq. (\ref{pichi}) it is possible to see the main obstruction to determinations of the hierarchy from near reactors.  The position of the $n$th peak is only sensitive to the mass combination
\beq
\Delta M^2_{eff}=\frac{\mn31}{1\pm\an/n}.
\eeq
As we will see, at low $n$, corresponding to peaks visible at relatively short baselines, the values of $\an$ are roughly linear.  Thus in this regime the positions of all of the peaks are sensitive to the same effective mass and so are independent of the hierarchy so long as that effective mass applies. {\textbf{The hierarchy can only be determined from the nonlinearity of $\an$.}}

\subsection{Finding $\an$}

The vector $\an$ encodes the effect of the neutrino mass matrix on the location of the survival probability peaks.  Therefore a knowledge of this function together with a measurement of the peaks allows one to reconstruct some elements of the mass matrix.

The values of $\an$ are determined from (\ref{pichi}) by the extrema of $P_{13}+P_{23}$
\bea
0&=&\frac{\partial}{\partial E}(P_{13}+P_{23})\propto \frac{\partial}{\partial E}\left[\cos\left(\frac{\mn31L}{2E}\right)+\tp212\cos\left(\frac{\mn32L}{2E}\right)\right]\\
&\propto&\sin\left(2\pi\an\right)+\left(1\pm\epsilon\right)\tp212\sin\left(2\pi[\an
\pm\epsilon(n+\an)]\right)\hsp
\epsilon=\frac{\m21}{\mn31}\nonumber
\eea
where the $-$ sign applies to the normal hierarchy and the $+$ sign to the inverse hierarchy.

As $\epsilon<<1$ and $\an<<n$ we may approximate
\beq
0\sim \sin(2\pi\an)+\tp212\sin(2\pi[\an\pm\epsilon n]). \label{alfeq}
\eeq
This can be expanded in a power series in $n$.  The highest energy peaks occur at small values of $n$, where the linear term in this expansion suffices
\beq
0\sim 2\pi(1+\tp212)\an\pm 2\pi\tp212\epsilon n
\eeq
which is easily solved for $\an$
\beq
\an\sim\mp\epsilon\sp212 n\sim \mp 0.011 n.
\eeq

Thus the n$th$ peak, for $n$ sufficiently small, lies at
\beq
\frac{L}{E}\sim\frac{4\pi}{\mn31}(1\mp\epsilon\sp212) n=\frac{4\pi n}{\mn31\pm\sp212\m21} \label{degen}
\eeq
where the lower sign corresponds to the normal hierarchy.

The basic problem facing shorter baseline experiments, which are only sensitive to peaks at small $n$, is already apparent in Eq. (\ref{degen}).  The mass difference $\mn31$ is degenerate with the hierarchy
\beq
\mn31(\mathrm{direct})= \mn31(\mathrm{inverse})+2\sp212\m21 .
\eeq
Therefore any experiment with such a short baseline that all observable peaks have $n$ in the regime in which $\alpha$ is linear are, alone, incapable of determining the hierarchy.  They can, however, determine the combination
\beq
\Delta M^2_{eff}=\mn31\pm\sp212\m21=\cp212\mn31+\sp212\mn32. \label{meff}
\eeq

Therefore reactor experiments can determine the hierarchy in only two ways.  Either an accurate determination of $\Delta M^2_{eff}$ can be combined with an accurate determination of another combination of the mass differences obtained from another experiment, a difference which needs to be known more precisely than the atmospheric mass difference measured by MINOS, or else the experiment needs to be sensitive to peaks at large enough $n$ that the linear approximation breaks down.  So just how large does $n$ need to be?

\subsection{Cubic terms is $\an$} \label{cubicosez}
To see where interference between 2-3 and 1-3 oscillations push the peaks of $P_{13}+P_{23}$ at larger values of $n$, we need to expand \an to cubic order
\beq
\an\sim  \mp\epsilon\sp212 n+b n^3. \label{cubico}
\eeq
and to substitute this expansion into Eq. (\ref{alfeq}).   The linear term in $\an$ already solves this equation at linear order.  At cubic order it yields
\beq
0\sim 2\pi(1+\tp212)b \pm\frac{4\pi^3}{3}\epsilon^3\sp212(\sp212-\cp212)
\eeq
and so
\beq
b\sim\mp\frac{2\pi^2}{3}\epsilon^3\sp212\cp212(\sp212-\cp212) \sim \pm 2\times 10^{-5} .
\eeq

The linear approximation to $\an$ is reliable when the cubic term in Eq. (\ref{cubico}) is much smaller than the linear term
\beq
n<<\sqrt{\frac{\epsilon\sp212}{b}}\sim 20.
\eeq
For example, at the 10th peak the contribution of the cubic term to the energy is only one quarter of that of the linear term.  The linear term we have seen shifts the effective mass by about 1\%, thus the cubic term, which is the leading hierarchy-dependent term, only shifts the energy of the tenth peak by about one quarter of a percent.  The other hierarchy would lead to a shift of a quarter percent in the other direction, so overall the difference in the energies of the 10th peaks in the normal and inverted hierarchy is about one half percent.

Therefore if only the first ten peaks can be measured at a given baseline, the detector will need to be able to determine the position of the tenth peak with a precision of a half percent for only a one sigma determination of the hierarchy, making a determination of the hierarchy using such an experiment alone quite unlikely.

\begin{figure} 
\begin{center}
\includegraphics[width=5in,height=2in]{alfa.pdf}
\caption{$\an$ for the normal hierarchy}
\label{anfig}
\end{center}
\end{figure}

The cubic expansion is no longer reliable for higher peaks, but $\an$ can be determined numerically.  As seen in Fig.~\ref{anfig} in the case of the normal hierarchy, it is periodic moduli $1/\epsilon$ and is zero at every multiple of $1/2\epsilon$.  But the main problem is that it is nearly linear for $n<10$, which is why the hierarchy is so nearly degenerate with the shift in the mass differences at all of these peaks.

A $2/k$ percent precision measurement of the peak energy of the the 14th peak would also give a $k$ sigma indication of the hierarchy.  The 14th peak can only be seen if it occurs at a high enough energy that the neutrino flux and detector resolution are sufficient to discern it.  For example if one requires it to appear at at least 3 MeV, so that the detector resolution may be better than 2\%, then using Eq. (\ref{corto}) the minimum baseline is about
\beq
L_{\mathrm{min}}\sim 45\ \mathrm{km}.
\eeq

\section{Determining mass differences from peak positions} \label{masssez}

\subsection{The reactor neutrino energy spectrum}

In the previous section we saw that the locations of the aperiodicity of the peaks is determined by the neutrino mass spectrum.  In particular, in $L/E$ space the peak positions are periodic modulo $1/\epsilon\sim 32$ and each set of $1/2\epsilon$ peaks is displaced about 1 percent either left or right depending on the hierarchy.  Thus, to determine the hierarchy, it suffices to measure the positions of enough peaks to within 2 percent precision.

The trouble with such a procedure is that no single experiment has access to all of the peaks, as one effectively has access to only a single baseline per detector and the energy spectrum is limited to that which is produced by a nuclear reactor, which effectively leads to a maximum usable neutrino energy.  Not even all of these energies are accessible as each type of neutrino detector has a minimum energy which it is able to detect.   To determine which peaks may be seen by a particular experiment, one must combine both of these constraints.

The neutrino flux from a reactor results almost entirely from decays of just 4 isotopes: \u35,\ \pu39,\ \pu41\ and \u38.  The flux $\phi_i(E)$ of neutrinos from each isotope $i$ is traditionally approximated as the exponential of a polynomial in the neutrino energy $E$~\cite{vogelengel}
\beq
\phi_i(E)=\mathrm{exp}\left(\sum_{k=1}^{m} a_{ki}E^{k-1}\right) \label{polinom}
\eeq
Theoretical errors on these fluxes are often claimed to be near the 2-3\% level, although recent theoretical fluxes~\cite{nuovoflusso} appear to be about 6\% higher than the fluxes measured at very short baseline experiments~\cite{reattoreanom} and at 1 kilometer experiments~\cite{noiunokm}.

The precision of such a phenomenological law depend on the degree $n$ of the fit polynomial.  For neutrinos with between 2 and 7.5 MeV, which will be the ones of interest in this note, it was shown in Ref.~\cite{huber2004} that a quadratic ($m=3$) fit tends to introduce errors of order 2-3\% whereas a 6 parameter ($m=6$) fit introduces errors of order 1\%, well below the error in the theoretical flux.  The differences between these parameterizations are oscillations over a characteristic scale of 1-2 MeV, which are likely too broad to give a false signal for a 1-3 oscillation peak, but may well disguise the depth of 1-2 oscillations and so affect the measured value of $\theta_{12}$.

The most recent theoretical estimate of the flux is Fig 53 of Ref.~\cite{cartabianca}, which shows a systematic 3\% excess over last year's estimates~\cite{nuovoflusso} at energies about 6 MeV and a one percent deficit beyond 4 MeV, which is well within the theoretical errors of the calculations.  This correction again will affect a single detector determination of $\theta_{12}$.

Not all of the neutrinos which are generated by the reactor will be measured.   The maximum number of neutrinos which can be measured at a given energy $E$ is the product of the produced flux with the fraction of neutrinos at that energy which can be measured.  For example, if neutrinos are measured via the inverse $\beta$ decay reaction
\beq
\overline{\nu}_e+p\rightarrow n+e^+
\eeq
then the maximum number of neutrinos detected is the flux/area at the baseline $L$ multiplied by the inverse $\beta$ decay cross section which at tree level is~\cite{sezionedurto}
\beq
\sigma(E)=0.0952\times 10^{-42}\mathrm{cm}^2(E_e p_e/\mathrm{MeV}^2) \label{albero}
\eeq
where the positron energy and momenta are, ignoring the neutron recoil, given in terms of the neutron, proton and electron rest masses
\beq
E_e=E-m_n+m_p+m_e\sim E-780\ \mathrm{keV}\hsp
p_e=\sqrt{E_e^2-m_e^2}. \label{posenergia}
\eeq

\begin{figure} 
\begin{center}
\includegraphics[width=5in,height=2in]{Spettro_NO.pdf}
\caption{The theoretical reactor neutrino spectrum as measured with inverse beta decay.}
\label{flussoorig}
\end{center}
\end{figure}

Combining the reactor flux (\ref{polinom}), with the coefficients of Ref.~\cite{vogelengel}, with the tree level cross section (\ref{albero}), one obtains the theoretical reactor flux, depicted in Fig.~\ref{flussoorig}.  While inverse $\beta$ decay is kinematically forbidden if $E<m_n+m_e-m_p\sim 1.8$\ MeV, it can be seen in Fig.~\ref{flussoorig} that the flux/energy is maximized at 3.6 MeV, falls to one third of its maximum by 6 MeV and to 10\% of its maximum by 7.5 MeV.  Thus useful information can only be obtained about the spectrum for energies within a factor of 2, which for each detector corresponds to values of $L/E$ within about a factor of 2.

\subsection{Effective masses at various baselines}

When neutrino oscillations are included, the theoretical flux $\Phi(E)$ from a reactor at a distance $L$ is then multiplied by the survival probability $P_{ee}$ given in Eq. (\ref{pee})
\beq
\Phi(E)=\sum_i b_i \phi_i(E)\sigma(E) P_{ee}(L/E)
\eeq
where $c_i$ is the quantity of each isotope in the reactor.

Using $\m21$ and $\spp2212$ from Ref.~\cite{gando}, $\mn32$ determined by combining neutrino and antineutrino mass differences from Ref.\cite{minosneut2012}  and $\spp2213$ from~\cite{noiunokm}
\beq
\m21=7.50\times 10^{-5}\hspp
\mn32=2.41\times 10^{-3}\hspp
\spp2212=0.857\hspp
\spp2213=0.096
\eeq
where $\mn31$ is determined using the normal and inverted hierarchies, this total neutrino flux is given in Fig.~\ref{tuttiflussi} at baselines of 40, 50, 60 and 70 km.

\begin{figure} 
\begin{center}
\includegraphics[width=3.2in,height=2in]{Plot_40_no_shift.pdf}
\includegraphics[width=3.2in,height=2in]{Plot_50_no_shift.pdf}
\includegraphics[width=3.2in,height=2in]{Plot_60_no_shift.pdf}
\includegraphics[width=3.2in,height=2in]{Plot_70_no_shift.pdf}
\caption{Theoretical neutrino fluxes, including 3 flavor oscillation, for both hierarchies as seen at 40, 50, 60 and 70 km.}
\label{tuttiflussi}
\end{center}
\end{figure}

The numbers $n$ of the local maxima can be read from (\ref{pichi}) by approximating $\mn31\sim\mn32$ and setting $\an=0$
\beq
n\sim\frac{4\pi}{\mn32}\frac{L}{E}\sim 0.9 \frac{L/\textrm{km}}{E/\textrm{MeV}}. \label{nvalore}
\eeq
As the error on $\mn32$ is of order 3\% and the difference between $\Delta M^2_{eff}$ and $\mn32$ is also of order 2\%, one may expect an error of 3-5\% in Eq. (\ref{nvalore}).  The fractional energy difference between the $n$th and $(n+1)$st peak is $1/n$.  Therefore an optimal detector can determine $n$ given a single peak only if $n$ is less than about 20, whereas an ordinary detector, ideally calibrated by radioactive decays close to the energy in question, can reliably determine the value $n$ of a peak when $n\leq 10$.

Although the values of peaks in Fig~\ref{tuttiflussi} are strongly dependent upon the hierarchy at every baseline, this does not mean that a measurement of the spectra can actually allow one to determine the hierarchy.  The problem, as was described in Sec.~\ref{teorsez}, is that the energies of the first 10 or so peaks only determine the mass difference $\Delta M^2_{eff}$ of Eq (\ref{meff}).

\begin{figure} 
\begin{center}
\includegraphics[width=3.2in,height=2in]{Plot_40_sovrapposto.pdf}
\includegraphics[width=3.2in,height=2in]{Plot_58_sovrapposto.pdf}
\caption{Theoretical neutrino fluxes, including 3 flavor oscillation, at 40 and 58 km for both hierarchies with the same value of $\Delta M^2_{eff}$.  In the left panel, at 40 km, the hierarchies are difficult to distinguish because $\an$ is nearly linear at the visible peaks.  This degeneracy is broken by the higher $n$ peaks visible at 58 km, seen in the right panel.}
\label{degenfig}
\end{center}
\end{figure}

Thus if the baseline is low enough so that only these peaks may be reliably measured, then there will be a value of $\Delta M^2_{eff}$ that reproduces the peaks for both hierarchies, as seen at 40 km in the first panel of Fig.~\ref{degenfig}.   Here the peak $n=5$ can barely be seen at 6.6\ MeV, whereas $n=6$ at 5.5 MeV is clearly discernible.  The largest peaks are $n=7,\ 8,\ 9$ amd $10$.  However the energies of these peaks are independent of the hierarchy at constant $\Delta M^2_{eff}$.  The hierarchy-dependence becomes somewhat larger at the 11th peak, which is located at about 3.35 MeV in the case of the normal hierarchy and 3.30 MeV in the case of the inverted hierarchy.  Thus if $\Delta M^2_{eff}$  peaks can be determined at better than 1\% from the low $n$ peaks and then the location of the 11th peak can be determined with a precision of better than 1\%, a determination of the hierarchy would be barely possible.  The lower energy peaks are more smaller, but more hierarchy dependent.  For example, the 15th peak would be at 2.50 MeV with the normal hierarchy, but 2.42 MeV with the inverted hierarchy.  This 3\% difference is well within the resolution of the proposed detectors of Refs. (\cite{caojun}), and so when combined with an accurate measurement of $\Delta M^2_{eff}$ one may could potentially determine the hierarchy at the 1-2$\sigma$ level.



In the second panel one can see the electron neutrino survival probability at 58 km with both hierarchies and the same value of $\Delta M^2_{eff}$.  At this long baseline one can see maxima up to $n=20$, where the nonlinearity in $\an$ is appreciable and so, as one can see, the low and mid energy peaks are hierarchy-dependent at the 1-2\% level.

\subsection{The 1-2 oscillation minimum}
It is clear from Fig.~\ref{degenfig} that if $\Delta M^2_{eff}$ is determined from the low $n$ peaks then the locations of the peaks at the 1-2 oscillation minimum
\beq
n=\frac{1}{2\epsilon}\sim 15\hsp
E=\frac{L}{18\mathrm{\ km}}\textrm{MeV}
\eeq
are hierarchy-dependent, and so one may hope to use their locations to determine the hierarchy.  This can be done if the peak locations allow for a combination of the mass differences which is distinct from $\Delta M^2_{eff}$.  In fact, two such combinations can be determined, one from the location of the peaks and one from the distance between them.

To derive these two combinations, we will need to find $\an$  near the 1-2 oscillation minimum.  This is easily obtained by expanding (\ref{alfeq}) about $n=1/2\epsilon$
\beq
0\sim \sin(2\pi\an)-\tp212\sin\left(2\pi\left[\an\pm\epsilon \left(n-\frac{1}{2\epsilon}\right)\right]\right)
\eeq
where again the positive sign applies to the inverted hierarchy.  Linearly expanding the sine function yields
\beq
0\sim (1-\tp212)\an\mp\epsilon \left(n-\frac{1}{2\epsilon}\right)
\eeq
and so for $n\sim 1/2\epsilon$
\beq
\an\sim\pm\epsilon \frac{n-\frac{1}{2\epsilon}}{ 1-\tp212}.  \label{anminimo}
\eeq

At $n=1/2\epsilon$, $\an=0$.  Of course $n$ must be an integer, so it can never be precisely equal to $1/2\epsilon$.  Nonetheless, $\an$ will be over order $0.01$ when $n$ is at the closest integral value, leading to a less than 0.1\% contribution to the energy.  When $\an=0$, the energy of the peak is
\beq
E=\frac{4\pi n L}{\mn31}=\frac{2\pi L}{\m21}
\eeq
allowing for a precise determination of $\m21$.  Of course, to know that one is at the 1-2 minimum by counting peaks, one must know $\epsilon$ and so this is related to a measurement of $\mn31$.  Alternately, one can find the 1-2 minimum by looking at the large oscillations in the flux due to 1-2 mixing and then count peaks to determine $n$ at the minimum, which allows for a determination of $\epsilon$ and so $\mn31$ from $\m21$.

The distance between the peaks near the 1-2 minimum allows for a determination of the mass differences which is independent of precise knowledge of the 1-2 oscillation parameters.  Inserting (\ref{anminimo}) in (\ref{pichi}) one finds that the distance between two peaks is
\beq
\Delta\left(\frac{L}{E}\right)=\frac{4\pi}{\mn31}\left(1\pm\frac{\epsilon}{1-\tp212}\right)
\eeq
and so it determines the effective mass
\beq
\Delta M^2_{min}=\left(1\mp\frac{\epsilon}{1-\tp212}\right)\mn31=\frac{2-\tp212}{1-\tp212}\mn31-\frac{1}{1-\tp212}\mn32 .
\eeq

This effective mass is quite different from that defined in Eq. (\ref{meff}).  Approximating $\tp212=1/2$ one finds that the spacing between the first 10 peaks yields an effective mass
\beq
\Delta M^2_{eff}=\frac{2}{3}\mn31+\frac{1}{3}\mn32 \label{meffdue}
\eeq
while the peaks between $n=14$ and $n=18$ yield
\beq
\Delta M^2_{min}=3\mn31-2\mn32 . \label{mmindue}
\eeq
An detector at a baseline of less than 50 km can accurately determine the combination (\ref{meffdue}) while one with a baseline between 45 and 70 km, as it sees higher $n$ peaks, may more easily measure the combination (\ref{mmindue}).   The normal mass hierarchy is equivalent to the second mass being larger than the first, and so by comparing these masses one can determine the hierarchy.

The difference between these masses is quite large, about 7\%, however a 7\% measurement of $\Delta M^2_{min}$ requires a 7\% precision in the measurement of the difference between the the peaks in the linear regime near to the 1-2 oscillation minimum $n=1/2\epsilon$.   As can be seen in Fig~\ref{anfig}, this linear regime includes about 8 peaks, corresponding to a 40\% variation in energy.  Therefore a 7\% precision in the distance between the peaks requires a 4\% precision in the energy of the peaks, much less than was required at shorter baselines.  Therefore a comparison of the low $n$ peak positions and the 1-2 minimum peak separations is a promising test of the hierarchy, so long as the statistics are sufficient for the peaks to be observed.

\section{Resolution and flux constraints} \label{vincsez}

\subsection{Energy resolution}

The energy of the neutrino $E$ is determined from the positron energy $E_e$ by subtracting 780 keV (\ref{posenergia}).   The positron energy is determined by counting photoelectrons in a scintillator.  The number of photoelectrons is proportional to the positron energy, and so the energy resolution $\sigma_E$ is proportional to the square root of the positron energy.  For example, at Daya Bay II it has been suggested in Ref.~\cite{caojun} that a resolution of
\beq
\sigma_E=0.03\sqrt{(E_e){\mathrm{MeV}}}.
\eeq
The observed positron  energy spectrum is then the convolution of the true spectrum with a Gaussian of width $\sigma$
\beq
P(E_e^{\textrm{observed}})=\int dE_e P(E_e) e^{-\frac{(E_e-E_e^{\textrm{observed}})^2}{2\sigma_E^2}}.
\eeq
Both spectra are displayed in Fig.~\ref{smearfig}.

\begin{figure} 
\begin{center}
\includegraphics[width=3.2in,height=2in]{Smear_40.pdf}
\includegraphics[width=3.2in,height=2in]{Smear_50.pdf}
\includegraphics[width=3.2in,height=2in]{Smear_60.pdf}
\includegraphics[width=3.2in,height=2in]{Smear_70.pdf}
\caption{The true neutrino spectrum and the measured spectrum with a resolution of $3\%/E$ using the normal hierarchy at baselines of 40, 50, 60 and 70 km.  Notice that the lowest energy oscillations are smeared away and the energy threshold for this smearing increases with the baseline, such that at higher baselines the maximum $n$ observable increases slowly.}
\label{smearfig}
\end{center}
\end{figure}

Even the observed spectrum is never observed.  It is the best that can be hoped for, once the response of the detector is understood and with an infinite number of neutrinos.  Finite flux effects will be briefly discussed in Subsec.~\ref{flussosez} and then discussed in detail in the companion paper on our simulation results.  A poorly understood detector response is fatal to the Fourier analysis that will be discussed in Sec.~\ref{fouriersez}, but individual peaks may be analyzed where the response is understood.

Even in this idealized setting, it is clear from Fig.~\ref{smearfig} that the low energy (high $n$) peaks cannot be resolved.  How high is $n$ for the biggest peak that can be resolved?  This depends on the baseline and the neutrino flux.  However a rough answer is obtained by asserting that the distance between the $n$th maximum and the adjacent minimum (\ref{corto})
\beq
\Delta E=0.9\frac{L/\mathrm{km}}{n}\mathrm{MeV}-0.9\frac{L/\mathrm{km}}{n+1/2}\mathrm{MeV}= 0.45 \frac{L/\mathrm{km}}{n^2}\mathrm{MeV}
\eeq
be greater than
\beq
2\sigma =0.06\sqrt{(E_e){\mathrm{MeV}}}=0.06\mathrm{\ MeV}\sqrt{\frac{L/\mathrm{km}}{n}-0.8}.
\eeq
These two quantities are equal when
\beq
L/\mathrm{km}=8\times 10^{-3}n^3\left(1\pm\sqrt{1-\frac{225}{n^2}}\right). \label{lnecc}
\eeq
The equality with the positive sign yields the upper bound on observable $n$ for a given baseline $L$, whereas the other typically yields a value of $n$ at energies where no reactor neutrinos are observed.

Alternately (\ref{lnecc}) provides the baseline $L$ necessary to observe the $n$th peak.  When $n\leq 15$ it provides no bound at all, so long as there is enough flux and the detector response is well understood, the energy resolution is not an obstruction for observing these peaks.  Of course at low baselines they may be difficult to observe because there simply are not many or any neutrinos observed at the corresponding energy.  At $n=16$, $17$ and $18$\ one finds minimum baselines of 45 km, 58 km and 72 km respectively.  Of course the true minimum depends on the neutrino flux.  But this rough estimate shows an essential point, that with a $3\%/\sqrt{E}$ fractional resolution any medium baseline, between about 40 and 70 km, is sufficient to observe the 1-2 oscillation minimum if there is enough neutrino flux.  Longer baselines only marginally extend the reach to higher peaks, although each of these peaks is in the 1-2 minimum region and so even just 1 or 2 more peaks can greatly enhance the possibility of determining the mass hierarchy.

While maximum observable $n$ is reasonably independent of the baseline, this derivation shows that it is strongly dependent upon $\sigma_E$.  A resolution comparable to that of Daya Bay or RENO would lead to a maximum $n$ which is still in the linear region of $\an$, and so the hierarchy would be unobservable.

\subsection{How much neutrino flux is required?} \label{flussosez}

The neutrino flux that can be observed by a large, distant detector via inverse beta decay is still quite unknown, even the efficiency of the detector is difficult to predict.  A rough estimate can be made using the flux observed at Daya Bay, using the flux normalization-independent determination of $\t13$ to eliminate the loss due to 1-3 operation.  At Daya Bay 17.4 GW of thermal power yields 80 neutrinos/day at each 20 ton detector at a weighted baseline of 1600 meters.  Therefore the total measured flux/year from a $P$ GW reactor complex measured at a detector of mass $M$ at a baseline $L$ is roughly
\beq
\Phi=365\times 80\times \frac{P}{17.4\ \mathrm{GW}}\frac{M}{0.02\ \mathrm{ktons}}\frac{2.6\ \mathrm{km}^2}{L^2}=2.2\times 10^5\frac{(P/\mathrm{GW})(M/\mathrm{ktons})}{(L/\mathrm{km})^2}. \label{totflusso}
\eeq
For example, at a distance $L$ from the Daya Bay and Ling Ao reactors a 20 kton detector may observe $7.6\times 10^7/L^2$ neutrinos/year, where the length $L$ is measured in kilometers.

Let the energy width of a peak be $\Delta E$, the integrated flux within that energy range be $\phi$ and the peak be a fraction $A$ higher than the flux in that energy range from the reactor in question after 1-2 oscillations plus all other reactors.   In the absence of other reactors the fraction $f$ varies from $\spp2213$ at low $n$ to $4\spp2213$ at the 1-2 minimum.  The fractional error in the flux in that range will be $1/\sqrt{\phi}$.  Therefore the peak may be observed if $\sqrt{\phi} A>1$.

Simply observing the peaks is useful for two reasons.  First of all, the distance between peaks is roughly $L\mn31/\pi$ and so by identifying peaks, one can check the consistency of the location of other peaks.  Second, recall that it is easier to determine the $n$ value of the well-separated, high energy, low $n$ peaks.  If one can observe the peaks from low to high $n$ then it is possible to count them and so determine the $n$ values of the low energy peaks.  While the hierarchy can be determined from the distance between the high $n$ peaks, as described in the previous subsection, without knowing the precise value of $n$, nonetheless if one knows the $n$ value of a high $n$ peak it can be used to determine a combination of the mass differences via (\ref{pichi}).  This determination has a precision of $1/n$, and so at high $n$ it leads to a precise determination.

However, to determine the hierarchy it is not enough to observe the peaks, one must determine their energies as precisely as possible.  How precisely may they be determined?  They may be determined within an energy $\delta E$ if the neutrino surplus can be seen in width $\delta E$ bands within the peak.  This requires
\beq
1<\sqrt{\phi}\frac{A\delta E}{\Delta E}=
\frac{\sqrt{\phi}Af}{n}
\eeq
where $f$ is the fractional energy precision desired.  This implies that the maximal fractional precision with which the energy of a peak can be measured by a detector with perfect resolution is
\beq
f>\frac{1}{An\sqrt{\phi}}.
\eeq
As mentioned above, with no unwanted backgrounds from other reactors, $A$ varies between $0.1$ at the 1-2 maxima to $0.4$ at the minima.

Consider for example the tenth peak at a 40 km baseline, which lies at 3.6 MeV.  At this point $A\sim 0.2$.  The flux within the peak is about 5\% of the total flux (\ref{totflusso}), which each year at a 20 kton detector at 40 km from Daya Bay may be
\beq
\phi=0.05\times 7.6\times 10^7/(40)^2=2.4\times 10^3.
\eeq
Therefore, after $m$ years, the best resolution of an ideal detector would be
\beq
f_{min}=\frac{1}{0.2\times 10\sqrt{2.4m\times 10^3}}=\frac{1}{100\sqrt{m}}.
\eeq
Therefore an ideal detector can find the tenth peak energy to within $1/\sqrt{m}$ percent after 10 years.  As the peak width is greater than the resolution of Daya Bay or RENO, the high energy resolution proposed at a new medium baseline detector experiment in Ref.~\cite{caojun} will not significantly alter the determination of this peak.  Therefore one may expect $\Delta M^2_{eff}$ to be determined to a precision significantly better than 1 percent at a 40 kilometer baseline experiment with no backgrounds from other reactors.

However, to determine the hierarchy, one also needs to determine another combination of the neutrino mass differences.  This requires the measurement of a higher $n$ peak.  Consider for example the $n=15$ peak at 2.5 MeV.  While it is in general a poor approximation to ignore the backgrounds provided by distant reactors, if one does ignore them then since since peak is near the 1-2 minimum, the relative peak height is $A=0.4$.   The total flux in the peak is only about 0.5
\beq
f_{min}=\frac{1}{0.4\times 15\sqrt{2.4m\times 10^2}}=\frac{1}{90\sqrt{m}}.
\eeq
As can be seen in Fig~\ref{degenfig}, a 3\% precision required to determine the hierarchy.  As the width of the peak is about 3\%, a $3\%/\sqrt{E}$ resolution may reduce the amplitude by a factor of 2, still allowing for a determination of the hierarchy.

\section{The Fourier transform of the survival probability} \label{fouriersez}

While each peak provides some information regarding a combination of neutrino mass differences and therefore the hierarchy, it may well be that the fluxes are too weak or the backgrounds too small for the peaks to be reasonably well identified.  Complimentary information can be combined by combining the peaks.  As the electron survival probability is a sum of periodic cosine functions, they can be combined by a Fourier transform.  Even when individual peaks are hard to identify, the combination probed by the Fourier transform may well be visible and so may provide the best chance for determining the hierarchy~\cite{hawaii}.

\subsection{The complex Fourier transform}

For simplicity we will approximate the observed electron antineutrino spectrum by a Gaussian distribution in $L/E$ space
\beq
\Phi\left(\frac{L}{E}\right)=e^{\left(\frac{L}{E}-L\langle\frac{1}{E}\rangle\right)^2/\sigma^2} \label{spec}
\eeq
where $\langle 1/E \rangle$ is the average $1/E$ of a neutrino which arrives.  While with only slightly more complicated equations the following could be avoided, we will make the crude approximation that (\ref{spec}) is the neutrino spectrum after 1-2 neutrino oscillations, and that the expectation value of the inverse energy is therefore taken with respect to the 1-2 oscillated spectrum, which depends upon $L$.

1-3 oscillations affect this spectrum by introducing a modulation equal to $\Phi(L/E) P_{13}(L/E)$.  The Fourier transform of this modulation is
\bea
F_{13}(k)&=&\int d\left(\frac{L}{E}\right) \Phi(L/E)  P_{13}(L/E) e^{i\frac{kL}{E}}\\&=&\frac{\sigma\sqrt{\pi}\cp212}{4}\left(e^{\frac{\sigma^2}{4}\left(k+\frac{\mn31}{2}\right)^2}e^{i\left(k+\frac{\mn31}{2}\right)L\langle\frac{1}{E}\rangle}+(e^{\frac{\sigma^2}{4}\left(k.\frac{\mn31}{2}\right)^2}e^{i\left(k-\frac{\mn31}{2}\right)L\langle\frac{1}{E}\rangle}\right)\nonumber
\eea
where we have factored the $\spp2213$ out of the definition of $F_{13}$. This quantity has two peaks, one at $k=\mn31/2$ and one at $k=-\mn31/2$.  At each peak, one of the two terms in the parenthesis dominates, and the other is suppressed by a factor of order $e^{n^2/4}$, which is large enough that we will neglect the subdominant term.  Therefore, near the first peak
\beq
F_{13}(k)=\frac{\sigma\sqrt{\pi}\cp212}{4}e^{\frac{\sigma^2}{4}\left(k-\frac{\mn31}{2}\right)^2}e^{i\left(k-\frac{\mn31}{2}\right)L\langle\frac{1}{E}\rangle}
\eeq
The complex norm of $F$ has a maximum at $k=\mn31/2$, where $F$ is real.  The real part of $F$, corresponding to a cosine transform, is, within the validity of the approximations described above, symmetric about this maximum.  The imaginary part, corresponding to a sine transform, vanishes at this maximum and is antisymmetric about it.

The transform $F_{23}(k)$ can be calculated identically.  Near the positive $k$ maximum the sum is just
\bea
F(k)&=&\frac{\sigma\sqrt{\pi}}{4}\left[\cp212 e^{\frac{\sigma^2}{4}\left(k-\frac{\mn31}{2}\right)^2}e^{i\left(k-\frac{\mn31}{2}\right)L\langle\frac{1}{E}\rangle}+\sp212 e^{\frac{\sigma^2}{4}\left(k-\frac{\mn32}{2}\right)^2}e^{i\left(k-\frac{\mn32}{2}\right)L\langle\frac{1}{E}\rangle}\right]\nonumber\\
&=&\frac{\sigma\sqrt{\pi}}{4}\left[\cp212 e^{\frac{\sigma^2}{4}\left(k-\frac{\mn31}{2}\right)^2}+\sp212 e^{\frac{\sigma^2}{4}\left(k-\frac{\mn32}{2}\right)^2}e^{\pm i\frac{\m21 L}{2}\langle\frac{1}{E}\rangle}\right]e^{i\left(k-\frac{\mn31}{2}\right)L\langle\frac{1}{E}\rangle} \label{xform}
\eea
where the $+$ sign applies to the normal hierarchy.

The first term corresponds to the old peak, at $k=\mn31$ and the second to a new peak, which on its own would be $\tp212\sim 1/2$ as high as the first, as $k=\mn32$.  Of course, due to the phase difference of the two terms, for some choice of parameters the interference between the two terms implies that the second is not a local maximum of the norm of the Fourier transform.  However, each term is maximized when it is real, and so each term can be seen as a peak or a shoulder in the Fourier cosine transform.  The $k$ value of the peak then gives the corresponding mass difference.  Thus if the smaller peak is to the left, as smaller $k$, of the larger peak then $\mn32<\mn31$ and so one can conclude that there is a normal neutrino mass hierarchy~\cite{hawaii}.

\subsection{Determining $\mn31$ at the 1-2 oscillation minimum}

We will refer to the baseline
\beq
L=\frac{2\pi}{\m21\langle\frac{1}{E}\rangle}
\eeq
as the 1-2 oscillation minimum, as it is roughly the baseline at which the largest fraction of neutrinos disappears as a result of $P_{12}$.   It is approximately 58 km.  At this distance the Fourier transform (\ref{xform}) simplifies as the two terms are precisely out of phase
\beq
F(k)=\frac{\sigma\sqrt{\pi}}{4}\left[\cp212 e^{\frac{\sigma^2}{4}\left(k-\frac{\mn31}{2}\right)^2}-\sp212 e^{\frac{\sigma^2}{4}\left(k-\frac{\mn32}{2}\right)^2}\right]e^{i\left(k-\frac{\mn31}{2}\right)\frac{2\pi}{\m21}} . \label{segno}
\eeq

The cosine transform is just the real part of $F$.  Up to $k$-independent factors, it is proportional to
\beq
F_{\cos}(\tilde{k})=\left( e^{-\frac{\sigma^2}{4}\tilde{k}^2}-\tp212  e^{-\frac{\sigma^2}{4}\left(\tilde{k}\pm\frac{\m12}{2}\right)^2}\right)\cos\left(\frac{2\pi\tilde{k}}{\m21}\right) \label{coseq}
\eeq
where we have defined the distance to the $1-3$ peak to be
\beq
\tilde{k}=k-\frac{\mn31}{2}.
\eeq
The maxima of $F_{\cos}$ are found by setting to zero its derivative with respect to $\tilde{k}$, yielding the condition
\bea
&&\left(\frac{\sigma^2}{2}\tilde{k}+\frac{2\pi}{\m21}\tan\left(\frac{2\pi\tilde{k}}{\m21}\right)\right)e^{-\frac{\sigma^2}{4}\tilde{k}^2}\\&&\ \ \ \ \ \ \ \ \ \ \ \ \ =
\tp212\left(\frac{\sigma^2}{2}\left(\tilde{k}\pm\frac{\m21}{2}\right)+\frac{2\pi}{\m21}\tan\left(\frac{2\pi\tilde{k}}{\m21}\right)\right)e^{-\frac{\sigma^2}{4}\left(\tilde{k}\pm\frac{\m21}{2}\right)^2}\nonumber
\eea
where again the $+$ sign corresponds to the normal hierarchy and the $-$ sign to the inverted hierarchy.  The largest peak is the closest to $\tilde{k}=0$, the absolute maximum of the Fourier transform of $P_{13}$.  To find this peak we may expand $\tilde{k}$ about 0, at linear order we find
\beq
\tilde{k}=\pm\frac{\m21/2}{1+\frac{\pi^2}{2\beta}\left(e^\beta\ctp212-1\right)+2\beta} \label{ktilde}
\eeq
where we have defined the constant
\beq
\beta=\frac{\sigma^2\m21}{16}.
\eeq

As the denominator of (\ref{ktilde}) is much greater than 1, we learn that
\beq
|\tilde{k}|<<\frac{\m21}{2}
\eeq
and so the maximum of $F_{\cos}$ lies at approximately
\beq
k_{max}=\frac{\mn31}{2}.
\eeq
This means that if the detector is placed at the 1-2 oscillation minimum baseline then the peak of the cosine transform of the full neutrino spectrum lies essentially at the peak of $P_{13}$ alone.  This is not because because the frequencies of the $P_{13}$ and $P_{23}$ terms are similar, but because the absolute maximum of the Fourier transform of $P_{13}$, which corresponds to its frequency, happens to coincide with one of the minima of the cosine transform of $P_{23}$, which is not its frequency.  {\textbf{Thus, at the 1-2 oscillation minimum, the absolute maximum of the sum of two cosine transforms is coincident with that of $P_{13}$ alone, allowing for a direct and precise determination of $\mn31$}}.  This may be useful on its own even if the hierarchy has already been determined by NOvA of T2K.

\subsection{Determining the hierarchy at the 1-2 oscillation minimum}

Can this simplification also yield information about the hierarchy?  A precise determination of $\mn31$ combined with MINOS' best fit for $\mn32$ could give a 1$\sigma$ answer to this question, within 5 years MINOS+ may improve this to 2$\sigma$.  But with enough flux the peak structure Fourier transform alone yields some information about the hierarchy.

\begin{figure} 
\begin{center}
\includegraphics[width=5.2in,height=2.5in]{FCT_58_D.pdf}
\caption{The cosine Fourier transform of $P_{12}$ (blue), $P_{23}$ (purple), $P_{13}$ (green) and $P_{ee}$ (yellow) at 58 km. {\textbf{This isn't quite true because our conventions for the P's are different, can we redo this figure with our conventions???}} Note that the $P_{13}$ and $P_{23}$ curves are just out of phase, so that the total extrema coincide with those of $P_{13}$.}
\label{cosfig}
\end{center}
\end{figure}

As can be seen in Fig.~\ref{cosfig}, Just as the relative sign in (\ref{segno}) implied that the global maximum of the cosine transform of the total spectrum is essentially coincident with that of $P_{13}$, it also implies that the minima just to its left and right are coincident
\beq
k_{\min}^L=\frac{\mn31-\m21}{2}\hsp
k_{\min}^R=\frac{\mn31+\m21}{2}.
\eeq
These are minima for the simple reason that the cosine on the right of (\ref{coseq}) is equal to $-1$.  Substituting these values of $k$ into (\ref{coseq}) we find the values of the cosine transforms at the two minima
\beq
F_{\cos}(k_{min}^L)=-\left(e^{-\beta}-\tp212 e^{-\beta(-1\pm 1)^2}\right)\hsp
F_{\cos}(k_{min}^R)=-\left(e^{-\beta}-\tp212 e^{-\beta(1\pm 1)^2}\right).
\eeq
In the case of the normal hierarchy, the positive sign means that the second term is larger on the left, meaning that the minimum on the right is deeper.  In the case of the inverted hierarchy these two depths are interchanged, and so the peak on the left is deeper.  This is the criterion for determining the hierarchy from the cosine transform which was proposed in Ref.~\cite{caojun}.

\begin{figure} 
\begin{center}
\includegraphics[width=5.2in,height=2.5in]{FST_58_D.pdf}
\caption{The sine Fourier transform of $P_{12}$ (blue), $P_{23}$ (purple), $P_{13}$ (green) and $P_{ee}$ (yellow) at 58 km. {\textbf{This isn't quite true because our conventions for the P's are different, can we redo this figure with our conventions???}} Note that the $P_{13}$ and $P_{23}$ curves are just out of phase, so that the total extrema coincide with those of $P_{13}$.}
\label{sinfig}
\end{center}
\end{figure}

A similar analysis can be applied to the imaginary part of $F(k)$, which is obtained via a sine transform.  As $F(\mn31/2)$ is real, the sine transform vanishes at the maximum of the global cosine transform.  The sine transform then has a maximum on the right and a minimum on the left.  The opposition of the phases of the Fourier transforms of $P_{13}$ and $P_{23}$ at the 1-2 oscillation minima again imply that these extrema of the sine transform of the full spectrum roughly coincide with the extrema of the sine transform of $P_{13}$ alone, as can be seen at 58 km in Fig.~\ref{sinfig}.  They simply correspond to the values of $k$ for which the phase in (\ref{segno}) is $\pm\pi/2$
\beq
k^L=\frac{\mn31}{2}-\frac{\m21}{4}\hsp
k^R=\frac{\mn31}{2}+\frac{\m21}{4}.
\eeq
Defining the sine transform so as to take the imaginary part of $F$ with the same normalization as for the cosine transform one then finds
\beq
F_{\sin}(k^L)=-i\left(e^{-\beta/4}-\tp212 e^{-\frac{\beta}{4}(2\mp 1)^2}\right)\hsp
F_{\sin}(k^R)=i\left(e^{-\beta/4}-\tp212 e^{-\frac{\beta}{4}(2\pm 1)^2}\right)
\eeq
where the top sign corresponds to the normal hierarchy.  In the case of the normal hierarchy the second term in the minimum on the left is larger and so $|F_{\sin}(k^L)|<|F_{\sin}(k^R)|$, whereas in the case of the inverted hierarchy the maximum on the right is larger than the minimum on the left.  Thus we have reproduced the correlation between the hierarchy and the relative sizes of these extrema observed in Ref.~\cite{caojun}.

Note that in the case of the normal hierarchy both extrema are more positive and in the case of the inverted hierarchy both are more negative, in other words {\textbf{near its maximum $F$ has a positive imaginary part in the case of a normal hierarchy and a negative imaginary part in the case of an inverted hierarchy}}.  This provides a new, third test which allows one to extrapolate the hierarchy from the Fourier transform of the survival probability.  Of course the optimal indicator of the hierarchy will be a weighted sum of these three tests, with weights that may be determined by series of simulations.  The ability to distinguish the hierarchies may also be improved by convoluting the observed spectrum with a slowly varying function of $(L/E)$ which weights neutrinos near the 1-2 minimum more heavily.  One can also use simulated data to determine which weighting functions are optimal for this task.

\subsection{Nonlinear Fourier transform}
The Fourier transform methods described above essentially work because the phase at the maximum is determined by the hierarchy, positive for the normal hierarchy and negative for the inverted hierarchy.  In the case of just $1-3$ oscillations, this phase is zero, since the corresponding oscillations are a cosine and the real part of the Fourier transform is determined by the cosine transform.  However when $P_{23}$ is included the peaks of the untransformed spectrum move according to Eq. (\ref{pichi}).  The distance between the untransformed peaks changes, which in general  moves the Fourier transformed peak.

Critically, the untransformed $P_{13}+P_{23}$ also loses its mod $2/\mn31$ periodicity, as $\an$ is not a linear function of $n$.  A given detector is only sensitive to some of the peaks, corresponding to a certain region in Fig.~\ref{anfig}.  Such regions are generally dominated by a domain in which $\an$ can be approximated not by a linear function, but by a linear function plus a constant offset.  This constant offset implies that the convolution of $P_{13}+P_{23}$ with the reactor neutrino spectrum is not of the form $\cos(\kappa L/E)$ but rather of the form $\cos(\kappa L/E+c)$ where the sign of $c$ is determined by the hierarchy.  This offset, $c$, leads to a translation in $L/E$ space which, after the Fourier transform, becomes a phase.  Thus the hierarchy determines an overall phase of the Fourier transform, which leads to the observable indicators described in the last section.

The Fourier transform method is robust since it sums multiple peaks together, and so it requires less neutrinos than a direct analysis of the positions of the peaks.  However it is inefficient because, as was just described, it works by approximating the function $\an$ to be affine in the energy range which is probed.  So one might ask if the performance would be increased by performing not an ordinary Fourier transform, but a Fourier transform with the nonlinearity of $\an$ built in.  The nonlinearity depends upon $\an$ which depends on the hierarchy and also weakly upon the neutrino mass matrix.  One may therefore attempt a nonlinear Fourier transform with both choices of hierarchy, and if desired, a weight function $g(L/E)$.  Such a nonlinear cosine transform is given by
\beq
F(k)=\int d\left(\frac{L}{E}\right)\Phi(L/E) P_{ee}(L/E) g(L/E) \cos\left(k\frac{L}{E}\pm 2\pi\alpha\left(\frac{k}{2\pi}\frac{L}{E}\right)\right)
\eeq
where $\alpha$ is a function which interpolates between the discrete values of $\an$ for the normal hierarchy and the positive sign corresponds to the inverted hierarchy.  This transform will add all of the peaks together with the same phase for the correct hierarchy whereas the peaks will be distorted by the other hierarchy.  Therefore the correct hierarchy can be determined from the fact that the corresponding nonlinear Fourier transform will have a larger absolute maximum.

\section{Interference effects} \label{intsez}

\subsection{Interference between reactors separated by 1 km} \label{unoproblema}

Consider two reactors of equal thermal power separated by a distance $D$.  If a detector is a distance $L>>D$ from the nearest and if the baseline makes an angle $\theta$ with respect to the line passing through both reactors, then the distance from the detector to the far reactor will be $L+d$ where $d=D\cos(\theta)$.

Adding the flux from both reactors, one finds that the 1-3 oscillations interfere
\beq
P_{13}\propto \cos\left(\frac{\mn31L}{2E}\right)+ \cos\left(\frac{\mn31(L+d)}{2E}\right)=2 \cos\left(\frac{\mn31d}{4E}\right)\cos\left(\frac{\mn31(L+d/2)}{2E}\right).
\eeq
The neutrinos from the two sources are not coherent, it is not the wavefunctions that add, but the probabilities.  And the result is that the amplitude of the oscillations at energy $E$ is suppressed by a factor of
\beq
\cos\left(\frac{\mn31d}{4E}\right)= \cos\left(3\frac{d/\mathrm{km}}{E/\mathrm{MeV}}\right).
\eeq
In particular they annihilate entirely when
\beq
\frac{d}{\mathrm{km}}=0.5 \frac{E}{\mathrm{MeV}}.
\eeq

Consider for example the pair of Daya Bay reactors and the two pairs of Ling Ao reactors which lie along a line at a distance of 0.8 and 1.3 km from the Daya Bay reactors.  The Daya Bay II reactor location suggested in Ref.~\cite{caojunseminario} is at an angle of 15 degrees with respect to a continuation of this line, yielding $d=$1.2\ km for the nearest and furthest reactor pair.  As a result 1-3 oscillations at 2.4 MeV completely cancel between these two reactor pairs, leaving only the oscillations of the middle reactor, and so effectively damping the oscillation amplitude by a factor of 3, which implies that an equally precise measurement of a peak at that energy requires 9 times as much flux as it would have without interference.

At the peak energy, 3.6 MeV the annihilation is not complete, but is reduced by a factor of $\cos(1)=0.6$.  Thus the total amplitude of the peak from all three reactor pairs is reduced by about 30\%, and so twice the neutrino flux will be necessary to observe these peaks.

This problem can be avoided if the detector is equidistant from all of the reactors.  In the case of the Daya Bay and Ling Ao complex, in the case of RENO and somewhat trivially in the case of Double Chooz this is possible as all of the reactors essentially line along a line.  It means however that such a detector will not be equidistant from any reactor that may eventually be built at Haifeng, thus reducing the neutrino fluxes assumed in Ref.~\cite{caojun} by a factor of 2.  Worse yet, the Haifeng reactor than provides a strong and undesirable background, as will be described in Subsec.~\ref{centoproblema}.

\subsection{Interference between reactors separated by 100 km} \label{centoproblema}

The 1-2 oscillation minimum
\beq
L=\frac{2\pi E}{\mn31}=16\left(\frac{E}{\mathrm{MeV}}\right)\mathrm{km}
\eeq
provides an ideal baseline to determine the mass hierarchy for a number of reasons.  Among these is that while the $P_{13}+P_{23}$ oscillation amplitude of
$(\cp212-\sp212)\spp2213$ is a factor of 3 less than its amplitude $\spp2213$ at the 1-2 maxima, the flux remaining after 1-2 oscillations is also smaller by a factor of $1/(1-\sp212\cp413)$ which is about 5.  Thus the 1-3 oscillations are a larger fraction of the total flux, making them easier to see above systematic errors, although not above statistical errors as $3>\sqrt{5}$.

However the smaller signal and smaller flux means that this part of the spectrum is particularly prone to interference from distant reactors, in particular those that may near the first 1-2 oscillation maximum
\beq
L=\frac{4\pi E}{\mn31}=32\left(\frac{E}{\mathrm{MeV}}\right)\mathrm{km}.
\eeq
In this case the addition factor of 5 in flux from the distant reactor outweighs the factor of 4 distance suppression, and so if both reactor complexes have the same strength than most of the flux at the 1-2 minimum is background.  This means that to obtain the same energy resolution at the 1-2 minimum peaks one will need more than 4 times as much neutrino flux.

This is the case for example with a detector placed 58 km away from the Daya Bay/Ling Ao complex, perpendicular to the reactors.  It would be at the 1-2 maximum of the proposed Haifeng reactors and would also suffer significant contamination from the Taishan and Yanjiang reactor complexes, at each of which at least 3 reactors are already under construction.  Similarly a reactor placed 60 km from Daya Bay, Ling Ao and Haifeng would be at the 1-2 maximum of the proposed 17.4 GW thermal power reactor complex at Lufeng.

As this effect increases the 1-2 oscillation minimum flux, it is also a serious obstacle to an accurate measurement of $\theta_{12}$.  If a model of the background neutrino flux from distant reactors is wrong, the error in the total flux can be compensated for by an error in the determination of $\theta_{12}$.  Ideally this problem can be solved by using two medium baseline detectors instead of one, which would break this degeneracy.  However in a world limited by costs, it may instead be necessary to simply try to make the background from other reactors as small as possible, thus minimizing the potential error in the determination of the hierarchy and in the determination of $\theta_{12}$.

Unlike the short distance interference problem discussed in Subsec.~\ref{unoproblema}, the fractional contamination caused by distance reactors depends on the baseline $L$ to the reactor whose neutrinos provide the signal.  The fractional contamination from undesired reactor neutrinos is inversely proportional to the desired signal strength, and so it is proportional to $L^2$.  Thus this problem is minimized by placing the detector as near to the reactor as possible.  If there is only one detector, and one wishes to use it to determine the hierarchy, then as explained in Subsec.~\ref{cubicosez}, this minimum distance cannot be shorter than about 45 km.

\section{Conclusions}

The newly discovered high value of $\theta_{13}$ means that the determination of the neutrino mass hierarchy at a medium baseline reactor experiment is now practical.  Previous analysis of such experiments have assumed values of $\spp2213$ an order of magnitude or more below its true value.  At such low values, individual peaks of the electron antineutrino spectra could not be resolved and one instead needed to rely upon a Fourier analysis~\cite{hawaii,caojun}.

In this note we have reconsidered the determination of the neutrino mass hierarchy now that 1-3 oscillations are large and so individual 1-3 peaks in the spectrum can easily be observed.  We found that the position of each peak determines a particular combination of the neutrino mass differences, for example the first 10 peaks all determine the same difference $\cp212\mn31+\sp212\mn32$.  In particular this means that the first 10 peaks alone cannot be used to determine the hierarchy, as they are fit equally well by the wrong hierarchy model in which this mass difference is preserved.  On the other hand we found that the positions of the next 10 peaks depend on distinct combinations of the mass differences, and so combining the energies at different peaks with some beyond the $10$th can lead to a determination of the hierarchy.  We also estimated the detector resolution and neutrino flux which are needed for such a goal, although an accurate determination will be left for the simulations to be discussed in our companion paper.

The information about the hierarchy is therefore contained in the low energy peaks, which suffer the most from the effects of poor resolution,  low neutrino flux and interference.  While the individual peaks are somewhat difficult to resolve, the situation can nonetheless be improved with a Fourier transform.  We have analyzed the Fourier transform of the spectrum, deriving the phenomenological hierarchy indicators proposed in Refs.~\cite{hawaii,caojun} and providing a new indicator, the complex phase at the peak of the Fourier transform, which we claim will be positive for the normal hierarchy and negative for the inverse hierarchy.  We also propose a new hierarchy-dependent nonlinear Fourier transform which will lead to a higher peak using the transform corresponding to the correct hierarchy.

The strength of the Fourier transform method is that it sums together all of the peaks to increase the strength of the signal with respect to the noise.  If the energy spectrum is shifted, this simply leads to an overall phase in the Fourier transform which does not seriously affect the analysis.  Likewise a uniform stretching of the spectrum simply shifts the peaks, which does not affect the determination of the hierarchy at all.  Any nonlinearity however poses a much more serious problem, as it can lead to an interference between the peaks in the spectrum which mutates or destroys the peak structure of the Fourier transform.  If the nature of the nonlinearity is known then one can adjust the analysis to correct for it~\cite{petcov2010} however if the nonlinearity is known it could be corrected directly in the reading of the energy.  In general the nonlinearity of the response is only known at energies where the detector has been calibrated with radioactive sources.  It may therefore be desirable to modify the Fourier transform so as to weight the more reliable energies more heavily.  One may also wish to weight the energies near the 1-2 minimum more heavily, as it contributes few neutrinos but it is indispensable in a determination of the hierarchy.  These  weights can be optimized by testing various hierarchy determination algorithms against simulated data.

We also discussed two interference effects.  First of all, reactors in the same array are generally separated by distances of order 1 km, which means that neutrinos arriving at a detector from one reactor at a 1-3 maximum may be at the same energy as those arriving from another at a 1-3 minimum.  The result is that the 1-3 oscillation signal can be severely reduced at the low energies in which this oscillation can occur within a reactor complex.  These are just the low energies which are necessary for the determination of the hierarchy, and so this interference poses a serious problem.  One solution is to place the detectors orthogonal to arrays of reactors.

The second interference effect arises from the fact that while the 1-2 oscillation minimum is the most useful energy range at which to determine the hierarchy, it enjoys a much lower neutrino flux than the 1-2 oscillation maximum.  As a result at these energy ranges one can expect serious contamination from distant reactors at their 1-2 maximum.  This problem, as well as the degeneracy between the neutrino flux from distant reactors and $\theta_{12}$, is optimally solved by using two detectors at different baselines.  However a cheaper solution to the same problem is to use a single detector at a shorter baseline, such as 45-50 km.


\section* {Acknowledgement}

\noindent
JE is supported by the Chinese Academy of Sciences
Fellowship for Young International Scientists grant number
2010Y2JA01. EC and XZ are supported in part by the NSF of
China.


\end{document}

\bibitem{lsnd}
A.~Aguilar-Arevalo {\it et al.}  [LSND Collaboration],
  ``Evidence for neutrino oscillations from the observation of anti-neutrino(electron) appearance in a anti-neutrino(muon) beam,''
  Phys.\ Rev.\ D {\bf 64} (2001) 112007
  [hep-ex/0104049].

\bibitem{minibooneanom}
A.~A.~Aguilar-Arevalo {\it et al.}  [The MiniBooNE Collaboration],
  ``A Search for electron neutrino appearance at the $\Delta m^{2} \sim 1$eV$^{2}$ scale,''
  Phys.\ Rev.\ Lett.\  {\bf 98} (2007) 231801
  [arXiv:0704.1500 [hep-ex]].
A.~A.~Aguilar-Arevalo {\it et al.}  [MiniBooNE Collaboration],
  ``Unexplained Excess of Electron-Like Events From a 1-GeV Neutrino Beam,''
  Phys.\ Rev.\ Lett.\  {\bf 102} (2009) 101802
  [arXiv:0812.2243 [hep-ex]].
 A.~A.~Aguilar-Arevalo {\it et al.}  [The MiniBooNE Collaboration],
  ``Event Excess in the MiniBooNE Search for $\bar \nu_\mu \rightarrow \bar \nu_e$ Oscillations,''
  Phys.\ Rev.\ Lett.\  {\bf 105} (2010) 181801
  [arXiv:1007.1150 [hep-ex]].

\bibitem{minosanom}
P.~Adamson {\it et al.}  [MINOS Collaboration],
  ``First direct observation of muon antineutrino disappearance,''
  Phys.\ Rev.\ Lett.\  {\bf 107} (2011) 021801
  [arXiv:1104.0344 [hep-ex]].

\bibitem{zichichi}
A.~Zichichi,
``Results from LVD-OPERA Combined Analysis: A Time-Shift in the OPERA Setup,"
available online at http://agenda.infn.it/getFile.py/access?resId=0\&materialId=slides\&confId=4896.

\bibitem{miniboonetuttobene}
E.~D.~Zimmerman [MiniBooNE Collaboration],
  ``Updated Search for Electron Antineutrino Appearance at MiniBooNE,''
  arXiv:1111.1375 [hep-ex].

\bibitem{minostuttobene}
P.~Adamson {\it et al.}  [MINOS Collaboration],
  ``An improved measurement of muon antineutrino disappearance in MINOS,''
  arXiv:1202.2772 [hep-ex].

\bibitem{nuovoflusso}
T.~.A.~Mueller, D.~Lhuillier, M.~Fallot, A.~Letourneau, S.~Cormon, M.~Fechner, L.~Giot and T.~Lasserre {\it et al.},
  ``Improved Predictions of Reactor Antineutrino Spectra,''
  Phys.\ Rev.\ C {\bf 83} (2011) 054615
  [arXiv:1101.2663 [hep-ex]].
P.~Huber,
  ``On the determination of anti-neutrino spectra from nuclear reactors,''
  Phys.\ Rev.\ C {\bf 84} (2011) 024617
   [Erratum-ibid.\ C {\bf 85} (2012) 029901]
  [arXiv:1106.0687 [hep-ph]].

\bibitem{reattoreanom}
G.~Mention, M.~Fechner, T.~.Lasserre, T.~.A.~Mueller, D.~Lhuillier, M.~Cribier and A.~Letourneau,
  ``The Reactor Antineutrino Anomaly,''
  Phys.\ Rev.\ D {\bf 83} (2011) 073006
  [arXiv:1101.2755 [hep-ex]].

\bibitem{smirnov}
P.~C.~de Holanda and A.~Y.~.Smirnov,
  ``Homestake result, sterile neutrinos and low-energy solar neutrino experiments,''
  Phys.\ Rev.\ D {\bf 69} (2004) 113002
  [hep-ph/0307266].
P.~C.~de Holanda and A.~Y.~.Smirnov,
  ``Solar neutrino spectrum, sterile neutrinos and additional radiation in the Universe,''
  Phys.\ Rev.\ D {\bf 83} (2011) 113011
  [arXiv:1012.5627 [hep-ph]].

\bibitem{icecube}
R.~Abbasi {\it et al.}  [IceCube Collaboration],
  ``Measurement of the atmospheric neutrino energy spectrum from 100 GeV to 400 TeV with IceCube,''
  Phys.\ Rev.\ D {\bf 83} (2011) 012001
  [arXiv:1010.3980 [astro-ph.HE]].

\bibitem{giuntireview}
  C.~Giunti and M.~Laveder,
  ``Implications of 3+1 Short-Baseline Neutrino Oscillations,''
  Phys.\ Lett.\ B {\bf 706} (2011) 200
  [arXiv:1111.1069 [hep-ph]].

\bibitem{sterilecosm}
J.~Hamann, S.~Hannestad, G.~G.~Raffelt and Y.~Y.~Y.~Wong,
  ``Sterile neutrinos with eV masses in cosmology: How disfavoured exactly?,''
  JCAP {\bf 1109} (2011) 034
  [arXiv:1108.4136 [astro-ph.CO]].

\bibitem{dayabay}
  F.~P.~An {\it et al.}  [DAYA-BAY Collaboration],
  ``Observation of electron-antineutrino disappearance at Daya Bay,''
  Phys.\ Rev.\ Lett.\  {\bf 108} (2012) 171803
  [arXiv:1203.1669 [hep-ex]].

\bibitem{piureattori}
L.~A.~Mikaelyan and V.~V.~Sinev,
  ``Neutrino oscillations at reactors: What next?,''
  Phys.\ Atom.\ Nucl.\  {\bf 63} (2000) 1002
   [Yad.\ Fiz.\  {\bf 63N6} (2000) 1077]
  [hep-ex/9908047].

\bibitem{neut2012}
D. Dwyer,
``Daya Bay Results," presented at Neutrino 2012 in Kyoto.
Soon to be available at http://neu2012.kek.jp/neu2012/programme.html.

\bibitem{doublechooz}
Y.~Abe {\it et al.}  [DOUBLE-Chooz Collaboration],
  ``Indication for the disappearance of reactor electron antineutrinos in the Double Chooz experiment,''
  Phys.\ Rev.\ Lett.\  {\bf 108} (2012) 131801
  [arXiv:1112.6353 [hep-ex]].

\bibitem{reno}
  J.~K.~Ahn {\it et al.}  [RENO Collaboration],
  ``Observation of Reactor Electron Antineutrino Disappearance in the RENO Experiment,''
  Phys.\ Rev.\ Lett.\  {\bf 108} (2012) 191802
  [arXiv:1204.0626 [hep-ex]].

\bibitem{nuturn}
``Observation of reactor neutrino disappearance at RENO," presented at $\nu$TURN 2012 under Gran Sasso.  Available at http://agenda.infn.it/contributionListDisplay.py?confId=4722.

\bibitem{globale1}
G.~L.~Fogli, E.~Lisi, A.~Marrone, A.~Palazzo and A.~M.~Rotunno,
  ``Evidence of $\theta_{13}$>0 from global neutrino data analysis,''
  Phys.\ Rev.\ D {\bf 84} (2011) 053007
  [arXiv:1106.6028 [hep-ph]].

\bibitem{globale2}
  T.~Schwetz, M.~Tortola and J.~W.~F.~Valle,
  ``Where we are on $\theta_{13}$: addendum to 'Global neutrino data and recent reactor fluxes: status of three-flavour oscillation parameters',''
  New J.\ Phys.\  {\bf 13} (2011) 109401
  [arXiv:1108.1376 [hep-ph]].

\bibitem{paloverde}
F.~Boehm, J.~Busenitz, B.~Cook, G.~Gratta, H.~Henrikson, J.~Kornis, D.~Lawrence and K.~B.~Lee {\it et al.},
  ``Final results from the Palo Verde neutrino oscillation experiment,''
  Phys.\ Rev.\ D {\bf 64} (2001) 112001
  [hep-ex/0107009].

\bibitem{chooz}
M.~Apollonio {\it et al.}  [Chooz Collaboration],
  ``Search for neutrino oscillations on a long baseline at the Chooz nuclear power station,''
  Eur.\ Phys.\ J.\ C {\bf 27} (2003) 331
  [hep-ex/0301017].

\bibitem{neutdarke}
X.~-J.~Bi, P.~-H.~Gu, X.~-l.~Wang and X.~-M.~Zhang,
  ``Thermal leptogenesis in a model with mass varying neutrinos,''
  Phys.\ Rev.\ D {\bf 69} (2004) 113007
  [hep-ph/0311022].
  R.~Takahashi and M.~Tanimoto,
  ``Model of mass varying neutrinos in SUSY,''
  Phys.\ Lett.\ B {\bf 633} (2006) 675
  [hep-ph/0507142].
  R.~Takahashi and M.~Tanimoto,
  ``Speed of sound in the mass varying neutrinos scenario,''
  JHEP {\bf 0605} (2006) 021
  [astro-ph/0601119].
  E.~Ciuffoli, J.~Evslin, J.~Liu and X.~Zhang,
  ``OPERA and a Neutrino Dark Energy Model,''
  arXiv:1109.6641 [hep-ph].

\bibitem{neal04}
  D.~B.~Kaplan, A.~E.~Nelson and N.~Weiner,
  ``Neutrino oscillations as a probe of dark energy,''
  Phys.\ Rev.\ Lett.\  {\bf 93} (2004) 091801
  [hep-ph/0401099].

\bibitem{tortola}
  M.~Tortola, J.~W.~F.~Valle and D.~Vanegas,
  ``Global status of neutrino oscillation parameters after recent reactor measurements,''
  arXiv:1205.4018 [hep-ph].

\bibitem{foglinuovo}
G.L. Fogli, E. Lisi, A. Marrone, D. Montanino, A. Palazzo and A.M. Rotunno,
``Global analysis of neutrino masses, mixings and phases: entering the era of leptonic CP violation searches,"
arXiv:1205.5254 [hep-ph].

\bibitem{bugey4}
Y.~Declais, H.~de Kerret, B.~Lefievre, M.~Obolensky, A.~Etenko, Y.~.Kozlov, I.~Machulin and V.~Martemyanov {\it et al.},
  ``Study of reactor anti-neutrino interaction with proton at Bugey nuclear power plant,''
  Phys.\ Lett.\ B {\bf 338} (1994) 383.

\bibitem{dayafeb}
F.~P.~An {\it et al.}  [Daya Bay Collaboration],
  ``A side-by-side comparison of Daya Bay antineutrino detectors,''
  arXiv:1202.6181 [physics.ins-det].

\bibitem{daya2007}
  X.~Guo {\it et al.}  [Daya-Bay Collaboration],
  ``A Precision measurement of the neutrino mixing angle theta(13) using reactor antineutrinos at Daya-Bay,''
  hep-ex/0701029.


\bibitem{tredueterm}
 A.~Melchiorri, O.~Mena, S.~Palomares-Ruiz, S.~Pascoli, A.~Slosar and M.~Sorel,
  ``Sterile Neutrinos in Light of Recent Cosmological and Oscillation Data: A Multi-Flavor Scheme Approach,''
  JCAP {\bf 0901} (2009) 036
  [arXiv:0810.5133 [hep-ph]].

\bibitem{neutrinoasym}
  S.~Hannestad, I.~Tamborra and T.~Tram,
  ``Thermalisation of light sterile neutrinos in the early universe,''
  arXiv:1204.5861 [astro-ph.CO].

\bibitem{baoscoperta}
  D.~J.~Eisenstein {\it et al.}  [SDSS Collaboration],
  ``Detection of the baryon acoustic peak in the large-scale correlation function of SDSS luminous red galaxies,''
  Astrophys.\ J.\  {\bf 633} (2005) 560
  [astro-ph/0501171].


\bibitem{bao}
W.~J.~Percival, S.~Cole, D.~J.~Eisenstein, R.~C.~Nichol, J.~A.~Peacock, A.~C.~Pope and A.~S.~Szalay,
  ``Measuring the Baryon Acoustic Oscillation scale using the SDSS and 2dFGRS,''
  Mon.\ Not.\ Roy.\ Astron.\ Soc.\  {\bf 381} (2007) 1053
  [arXiv:0705.3323 [astro-ph]].

\bibitem{cosmomc}
  A.~Lewis and S.~Bridle,
  ``Cosmological parameters from CMB and other data: A Monte Carlo approach,''
  Phys.\ Rev.\ D {\bf 66} (2002) 103511
  [astro-ph/0205436].

\bibitem{wmap}
E.~Komatsu {\it et al.}  [WMAP Collaboration],
  ``Seven-Year Wilkinson Microwave Anisotropy Probe (WMAP) Observations: Cosmological Interpretation,''
  Astrophys.\ J.\ Suppl.\  {\bf 192} (2011) 18
  [arXiv:1001.4538 [astro-ph.CO]].

\bibitem{Suzuki:2011hu}
  N.~Suzuki {\it et al.},
  ``The Hubble Space Telescope Cluster Supernova Survey: V. Improving the Dark
  Energy Constraints Above z>1 and Building an Early-Type-Hosted Supernova
  Sample,''
  arXiv:1105.3470 [astro-ph.CO].

\bibitem{h}
A.~G.~Riess, L.~Macri, S.~Casertano, H.~Lampeitl, H.~C.~Ferguson, A.~V.~Filippenko, S.~W.~Jha and W.~Li {\it et al.},
  ``A 3\% Solution: Determination of the Hubble Constant with the Hubble Space Telescope and Wide Field Camera 3,''
  Astrophys.\ J.\  {\bf 730} (2011) 119
   [Erratum-ibid.\  {\bf 732} (2011) 129]
  [arXiv:1103.2976 [astro-ph.CO]].

\bibitem{nessundivergenza}
J.~-Q.~Xia, G.~-B.~Zhao, B.~Feng, H.~Li and X.~Zhang,
  ``Observing dark energy dynamics with supernova, microwave background and galaxy clustering,''
  Phys.\ Rev.\ D {\bf 73} (2006) 063521
  [astro-ph/0511625].
G.~-B.~Zhao, J.~-Q.~Xia, M.~Li, B.~Feng and X.~Zhang,
  ``Perturbations of the quintom models of dark energy and the effects on observations,''
  Phys.\ Rev.\ D {\bf 72} (2005) 123515
  [astro-ph/0507482].

\bibitem{altroquintom}
  E.~Giusarma, M.~Archidiacono, R.~de Putter, A.~Melchiorri and O.~Mena,
  ``Sterile neutrino models and nonminimal cosmologies,''
  Phys.\ Rev.\ D {\bf 85} (2012) 083522
  [arXiv:1112.4661 [astro-ph.CO]].

\bibitem{unione2}
N.~Suzuki, D.~Rubin, C.~Lidman, G.~Aldering, R.~Amanullah, K.~Barbary, L.~F.~Barrientos and J.~Botyanszki {\it et al.},
  ``The Hubble Space Telescope Cluster Supernova Survey: V. Improving the Dark Energy Constraints Above z=1 and Building an Early-Type-Hosted Supernova Sample,''
  Astrophys.\ J.\  {\bf 746} (2012) 85
  [arXiv:1105.3470 [astro-ph.CO]].

\bibitem{quintom}
B.~Feng, M.~Li, Y.~-S.~Piao and X.~Zhang,
  ``Oscillating quintom and the recurrent universe,''
  Phys.\ Lett.\ B {\bf 634} (2006) 101
  [astro-ph/0407432].
  B.~Feng, X.~L.~Wang and X.~M.~Zhang,
``Dark energy constraints from the cosmic age and supernova,''
 Phys.\ Lett.\  B {\bf 607} (2005) 35
 [arXiv:astro-ph/0404224].
  X.~-F.~Zhang, H.~Li, Y.~-S.~Piao and X.~-M.~Zhang,
``Two-field models of dark energy with equation of state across -1,''
  Mod.\ Phys.\ Lett.\ A {\bf 21}, 231 (2006)
  [astro-ph/0501652].

\bibitem{neutrinoasymvecchio}
  R.~Foot and R.~R.~Volkas,
  ``Reconciling sterile neutrinos with big bang nucleosynthesis,''
  Phys.\ Rev.\ Lett.\  {\bf 75} (1995) 4350
  [hep-ph/9508275].

\bibitem{fr}
  H.~Motohashi, A.~A.~Starobinsky and J.~'i.~Yokoyama,
  ``Cosmology based on f(R) Gravity admits 1 eV Sterile Neutrinos,''
  arXiv:1203.6828 [astro-ph.CO].

\bibitem{robert}
R.~H.~Brandenberger, N.~Kaiser, D.~N.~Schramm and N.~Turok,
  ``Galaxy and Structure Formation with Hot Dark Matter and Cosmic Strings,''
  Phys.\ Rev.\ Lett.\  {\bf 59} (1987) 2371.
R.~H.~Brandenberger, A.~Mazumdar and M.~Yamaguchi,
  ``A Note on the robustness of the neutrino mass bounds from cosmology,''
  Phys.\ Rev.\ D {\bf 69} (2004) 081301
  [hep-ph/0401239].

\bibitem{monopoli}
J.~Evslin and S.~B.~Gudnason,
  ``High Q BPS Monopole Bags are Urchins,''
  arXiv:1111.3891 [hep-th].
J.~Evslin and S.~B.~Gudnason,
  ``Dwarf Galaxy Sized Monopoles as Dark Matter?,''
  arXiv:1202.0560 [astro-ph.CO].

\bibitem{lsndnonstandard}
  G.~Karagiorgi, M.~H.~Shaevitz and J.~M.~Conrad,
  ``Confronting the short-baseline oscillation anomalies with a single sterile neutrino and non-standard matter effects,''
  arXiv:1202.1024 [hep-ph].

\bibitem{envir}
  G.~Karagiorgi, M.~H.~Shaevitz and J.~M.~Conrad,
  ``Confronting the short-baseline oscillation anomalies with a single sterile neutrino and non-standard matter effects,''
  arXiv:1202.1024 [hep-ph].

\end{thebibliography}

\end{document}